\documentclass[twoside,11pt]{article}

\usepackage[accepted]{melba}

\usepackage{amsmath,amsfonts}

\usepackage{diagbox}
\usepackage{subcaption}
\usepackage{xcolor}
\usepackage[normalem]{ulem}

\melbaheading{2022:020}{https://www.melba-journal.org/papers/2022:020.html}{2022}{1-31}{02/2022}{07/2022}{Bastiaansen, Rousian, Steegers-Theunissen, Niessen, Koning and Klein}{Perinatal, Preterm and Paediatric Image Analysis (PIPPI) 2021 special issue}{Esra Abaci-Turk, Jana Hutter, Roxane Licandro, Christopher Macgowan, Andrew Melbourne}

\ShortHeadings{Multi-Atlas Segmentation and Spatial Alignment}{Bastiaansen et al.}
\firstpageno{1}

\title{Multi-Atlas Segmentation and Spatial Alignment of the Human Embryo in First Trimester 3D Ultrasound}

\author{\name Wietske A.P. Bastiaansen \email w.bastiaansen@erasmusmc.nl \\  
	\addr Biomedical Imaging Group Rotterdam, Department of Radiology and Nuclear Medicine, Erasmus MC,  University Medical Center Rotterdam, The Netherlands\\
	Department of Obstetrics and Gynaecology, Erasmus MC, University Medical Center Rotterdam, The Netherlands
	\AND
	\name Melek Rousian  \\
	\addr Department of Obstetrics and Gynaecology, Erasmus MC,  University Medical Center Rotterdam, The Netherlands
	\AND
	\name Régine P.M. Steegers-Theunissen \\
	\addr Department of Obstetrics and Gynaecology, Erasmus MC,  University Medical Center Rotterdam, The Netherlands
	\AND
	\name Wiro J. Niessen \\
	\addr  Biomedical Imaging Group Rotterdam, Department of Radiology and Nuclear Medicine, Erasmus MC,  University Medical Center Rotterdam, The Netherlands\\
	Faculty of Applied Sciences, Delft University of Technology, The Netherlands
	\AND
	\name Anton H.J. Koning \\
	\addr Generation R Study Group, Erasmus MC,  University Medical Center, Rotterdam, The Netherlands
	\AND
	\name Stefan Klein \\
	\addr Biomedical Imaging Group Rotterdam, Department of Radiology and Nuclear Medicine, Erasmus MC, Rotterdam,  University Medical Center, The Netherlands
}

\begin{document}

\maketitle

\begin{abstract}
Segmentation and spatial alignment of ultrasound imaging data acquired in the first trimester is crucial for monitoring early human embryonic growth and development throughout this important period of life. Current approaches are either manual or semi-automatic and are therefore very time-consuming and prone to errors. To automate these tasks, we propose a multi-atlas framework for automatic segmentation and spatial alignment of the embryo using deep learning with minimal supervision. Our framework learns to register the embryo to an atlas, which consists of the ultrasound images acquired at a range of gestational ages, segmented and spatially aligned to a predefined standard orientation. From this, we can derive the segmentation of the embryo and put the embryo in standard orientation. Ultrasound images acquired at 8+0 till 12+6 weeks gestational age were used and the ultrasound images of eight subjects were selected as atlas. We evaluated different fusion strategies to incorporate multiple atlases: 1) training the framework using atlases from a single subject, 2) training the framework with data of all available atlases and 3) ensembling of the frameworks trained per subject. To evaluate the performance, we calculated the Dice score over the test set. We found that training the framework using all available atlases outperformed ensembling and gave similar results compared to the best of all frameworks trained on a single subject. Furthermore, we found that selecting images from the four atlases closest in gestational age out of all available atlases, regardless of the individual quality, gave the best results with a median Dice score of $0.72$. We conclude that our framework can accurately segment and spatially align the embryo in first trimester 3D ultrasound images and is robust to the variation in quality that existed in the available atlases. Our code is publicly available at: \href{https://gitlab.com/radiology/prenatal-image-analysis/multi-atlas-seg-reg}{https://gitlab.com/radiology/prenatal-image-analysis/multi-atlas-seg-reg}.
\end{abstract}


\section{Introduction}
The periconception period, which runs from 14 weeks before conception till the end of the first trimester, is crucial for pre- and postnatal health \citep{SteegersTheunissen2013}.  Prenatal growth and development during this period is predominantly monitored via ultrasound imaging, mainly by manual measurements of the crown-rump length (CRL) and inspection of standard planes of the embryo  \citep{ISUOG}. Due to improvements in technology, and due to the introduction of trans-vaginal high frequency ultrasound probes, even more can be visualized already during the first trimester. This also let to the advise of the ISUOG, encouraging first trimester neuro-ultrasonography \citep{ISUOG2021}. 
Besides inspection of standard planes and the measurement of the CRL, nowadays due to 3D ultrasonography it is also possible to measure the volume of the embryo during the first trimester taking all dimensions of a human being into account, which provides more information than just one plane. However, these measurements are time consuming and prone to human errors, such as over- or underestimation of the volume due to manual delineation \citep{Carneiro2008,blaas2006}. Automating this process would lead to less investigation time and more consistent results.

Despite its challenges, 3D first trimester ultrasound is especially suitable for learning based methods, as the whole embryo, thanks to its limited size, can be imaged in one dataset. Here, we propose a deep learning based framework to automatically segment and spatially align the embryo to a standard orientation. These two tasks form the basis for automatic monitoring of growth and development: placing the embryo in a standard orientation enables derivation of the standard planes, and the segmentation of the embryo provides the embryonic volume. The advantage of the proposed framework is that after alignment, any standard plane and associated biometric measurements can easily be identified. Finally, real-time automatic alignment of the embryo simplifies the acquisition of standard planes in clinical practice, by providing probe movement guidance to visualize the plane \citep{droste2020}.

To achieve segmentation and spatial alignment simultaneously, we take an atlas-based registration approach using deep learning. We register all data to an atlas in which the embryo has been segmented and put in a standard orientation. To make our framework applicable to data acquired between the 8th and 12th week of pregnancy and to increase its robustness to variations in appearance of the embryo across pregnancies, we take a multi-atlas approach by using data of multiple subjects with longitudinally acquired ultrasound images during pregnancy. To address how to combine data from multiple subjects we compared three strategies, namely: 1) training the framework using atlases from a single subject, 2) training the framework with data of all available atlases and 3) ensembling of the frameworks trained per subject. 

By taking an atlas-based registration approach we circumvent the need for ground truth segmentations during training, as atlas-based registration can be fully unsupervised. Manual segmentations of the embryo are laborious to obtain and typically not available, since in clinical practice mainly the length of the embryo and various diameters are used \citep{ISUOG}. Due to the rapid development of the embryo in the first trimester and wide variation in spatial position and orientation of the embryo and presence of other structures such as placenta, umbilical cord and uterine wall, we choose to add supervision using landmarks to our framework. We proposed to use the crown and rump landmarks of the embryo as supervision during training, since the CRL measurement is relatively easy to obtain and  a standard measure in clinical practice \citep{Rousian2018}. During inference, the proposed method is fully automatic and does not rely on the crown and rump landmarks.

The main contributions of the research presented in this paper are:
\begin{enumerate}
    \item we propose the first automated method to segment and simultaneously spatially align the embryo in 3D ultrasound images acquired during the first trimester;
    \item we compare different strategies to incorporate data of multiple atlases and address the question how many atlas images to select;
    \item we circumvent the need for manual segmentations for training of our framework by relying only on minimal supervision based on two landmarks. 
\end{enumerate}

\section{Related work}
\subsection{Automatic analysis of first trimester ultrasound}
In literature there is little research available on automatic analysis of first trimester ultrasound in human pregnancies. Two studies focus on performing segmentation of the embryo in first trimester 3D ultrasound \citep{yang2018,looney2021}. Both approaches use a fully convolutional neural network, combined with transfer learning and either label refinement using a recurrent neural network \citep{yang2018} or a two-pathway architecture \citep{looney2021}. They achieve mean Dice score of 0.876 (104 3D volumes, 10-14 weeks GA) and 0.880 (2393 volumes, 11-14 weeks GA) respectively. \cite{Carneiro2008} studied automatic extraction of the biparietal diameter, head circumference, abdominal circumference, femur length, humerus length, and CRL in first trimester ultrasound using a constrained probabilistic boosting tree classifier. Finally, three studies focused on obtaining the head circumference using publicly available data, which consists of 2D ultrasound scans of the trans-thalamic plane  \citep{heuvel2018,al2019,li2020}.

Limitation of the aforementioned studies is that they focus on extracting individual measurements or segmentations, whereas we propose to align the embryo to a predefined standard space. This is beneficial since alignment to a standard space allows for straightforward extraction of various biometric and volumetric measurements in a unified framework.

\subsection{Joint segmentation and spatial alignment of prenatal ultrasound}
Most published studies focus on performing spatial alignment and segmentation separately; for a comprehensive overview see \cite{Torrents-barrena2019}. When both spatial alignment and segmentation are of interest performing them sequentially raises the question of the optimal order of operations. Especially since some studies focusing on performing segmentation require prior knowledge about spatial orientation of the embryo \citep{Gutierrez2013,Yaqub2013} and some studies focusing on standard plane detection require the segmentation or detection of structures within the embryo \citep{Ryou2016,Yaqub2015}. To circumvent this, a few studies perform both tasks at once \citep{Chen2012,Kuklisova-Murgasova2013,Namburete2018}. 

\cite{Chen2012} achieved both tasks for the fetal head using spatial features from eye detection and image registration techniques, for ultrasound images acquired at 19-22 weeks GA. However, the eye is not always clearly detectable in 3D first trimester ultrasound, making this method not applicable in our case.

\cite{Kuklisova-Murgasova2013} achieved both tasks for the fetal brain, using a magnetic resonance imaging (MRI) atlas and block matching using image registration techniques, for ultrasound images acquired at 23-28 weeks GA. However, such an atlas or reference MRI image is not available for the first trimester \citep{OISHI2019}.

\cite{Namburete2018} achieved both alignment and segmentation of the fetal head with a supervised multi-task deep learning approach, using slice-wise annotations containing a priori knowledge of the orientation of the head, and manual segmentations, for ultrasound images acquired at 22-30 weeks GA. However, this method assumes that during acquisition the ultrasound probe is positioned such that the head is imaged axially, which is not always the case for 3D first trimester ultrasound.

Although none of the studies are applicable to first trimester ultrasound, they all employ image registration techniques to perform segmentation and spatial alignment simultaneously. In line with this, we take an image registration approach and tailor our method to be applicable to first trimester ultrasound.

\subsection{Deep learning for image registration}
Recently, deep learning methods for image registration have been developed and show promising results in many areas; for an extensive overview see \cite{Boveiri2020}. 
The common assumption for most of the deep learning approaches for image registration is that the data is already affinely registered. However, due to the rapid development of the embryo and the wide variation in position and spatial orientation the affine registration is challenging to obtain and therefore has to be incorporated in the framework. 

There are a few studies focusing on achieving both affine and nonrigid registration. For example, \cite{DeVos2019}, proposed a multi-stage framework, dedicating the first network to learning the affine transformation and the second network to learning the voxelwise displacement field. These networks are trained in stages. \cite{Shen2019} proposed a similar approach where the main difference is that they used a diffeomorphic model for the nonrigid deformation. We also take a multi-stage approach with a dedicated network for the affine and nonrigid deformation. We based the design of both networks on Voxelmorph by \cite{Balakrishnan2018}, which is developed for deep learning-based nonrigid image registration and is publicly available.

\subsection{Multi-atlas segmentation with deep learning}
Using multiple atlases is a common and successful technique for biomedical image segmentation. Adding data of multiple atlases gives in general better and more robust results and is widely used in classical not learning-based methods \citep{Iglesias2015}. However, learning-based methods can also benefit from using multiple atlases \citep{Ding2019,Fang2019,Lee2019}. 

\cite{Fang2019} omits the registration step by directly guiding the training of a segmentation network using multiple atlases as ground truth. \cite{Ding2019} and \cite{Lee2019} both employ deep learning for the image registration step, but they use different strategies for training. \cite{Ding2019} trained a network that warps all available atlas to the target image at once. On the other hand, \cite{Lee2019} trains the network by giving at every epoch a random atlas to the network to register to the target image.

Both approaches train the network to be able to register all available atlases to all target images. In our case, the set of atlases consists of ultrasound images from different subjects acquired in every week GA. Due to the rapid development of the embryo, only atlases acquired at approximately the same GA as the image are useful for registration. Hence, we compared different fusion strategies that take this into account.

\subsection{Extension of preliminary results}
Preliminary results of this framework were published in \cite{Bastiaansen2020} and \cite{Bastiaansen2020a}. In \cite{Bastiaansen2020} we proposed a fully unsupervised atlas-based registration framework using deep learning for the alignment of the embryonic brain acquired at 9 weeks GA. We showed that having a separate network for the affine and nonrigid deformations improved the results. Subsequently, in \cite{Bastiaansen2020a} our previous work was extended by adding minimal supervision using the crown and rump landmarks to align and segment the embryo. 

Here, we substantially extend our previous work by making it applicable to data acquired between the 8th and 12th week of pregnancy. We achieve this by taking a multi-atlas approach with atlases consisting of longitudinally acquired ultrasound images from multiple pregnancies. We compare different fusion strategies to combine data from multiple subjects and address the influence of differences in image quality of the atlas.

Compared to our previous work, we performed a more in-depth hyperparameter search for our deep learning framework and improved the division of the data in training, validation and test set taking into account the distribution of GA and available information used for evaluation, such as embryonic volume. Furthermore, we compared our best model to the 3D nn-UNet \citep{isensee2021}, which is a current state-of-the-art fully supervised segmentation method. Finally, to obtain the segmentation of the embryo, in the proposed model the inverse of the nonrigid deformation is needed. To make our deformations invertible, we adopt a diffeomorphic deformation model as proposed by \cite{Dalca2019}, using the stationary velocity field.

\section{Method}
\subsection{Overview}
\begin{figure}[b!]
    \centering
    \includegraphics[width=\textwidth]{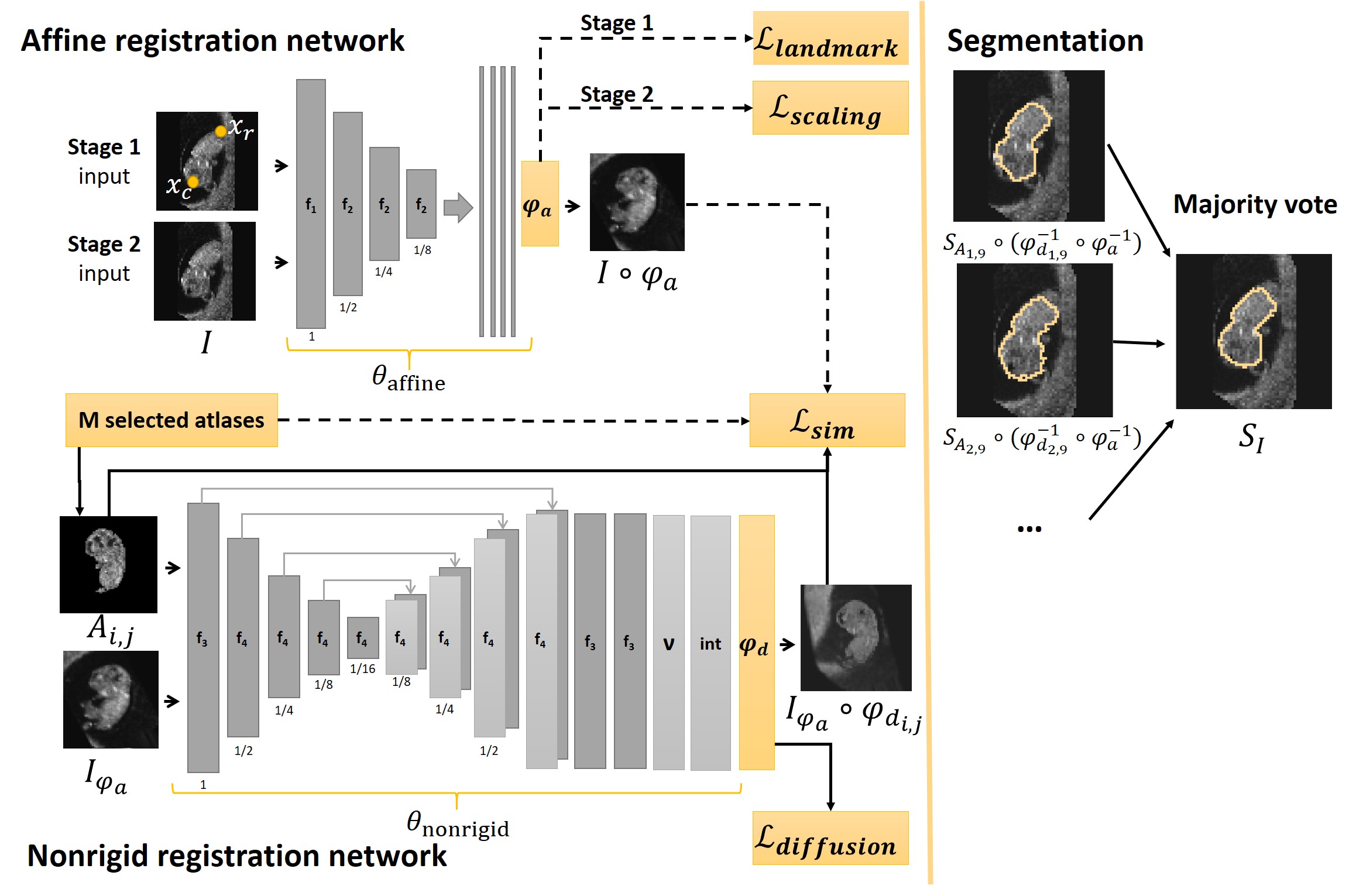}
    \caption{Overview of our framework to register image $I$ to atlas $A_{i,j}$. Atlas $A_{i,j}$ corresponds to subject $i$ with an ultrasound image taken in week $j$ of pregnancy. The network learns the affine and nonrigid deformations $\phi_a$ and $\phi_{d_{i,j}}$. Using the inverse deformations and majority voting we obtain the segmented images $S_I$. $I\circ\phi_a$ gives the image $I$ in standard orientation. }
    \label{fig:network_archi}
\end{figure}

We present a deep learning framework to simultaneously segment and spatially align the embryo in image $I$. We propose to achieve this by registering every image $I$ to an atlas via deformation $\phi$:
\begin{equation}A(x) \approx I \left(\phi(x)\right) \quad \forall x \in \Omega, \end{equation}
with $\Omega$ the 3D domain of $A,I$. We assume that the atlas $A$ is in a predefined standard orientation and the segmentation $S_A$ is available. Now, $I \circ \phi$ is in standard orientation and the segmentation of image $I$ is given by: 
\begin{equation} S_{I} := S_{A} \circ \phi^{-1}. \end{equation}
The deep learning framework learns to estimate deformation $\phi$ given atlas $A$ and the image $I$. We propose to use a multi-atlas approach by registering every image to the $M$ atlases that are closest in GA. In Fig. \!\ref{fig:network_archi} a detailed overview of our framework can be found; all components are explained in the remainder of this section. 
\begin{figure}[t!]
    \centering
    \includegraphics[width=\textwidth]{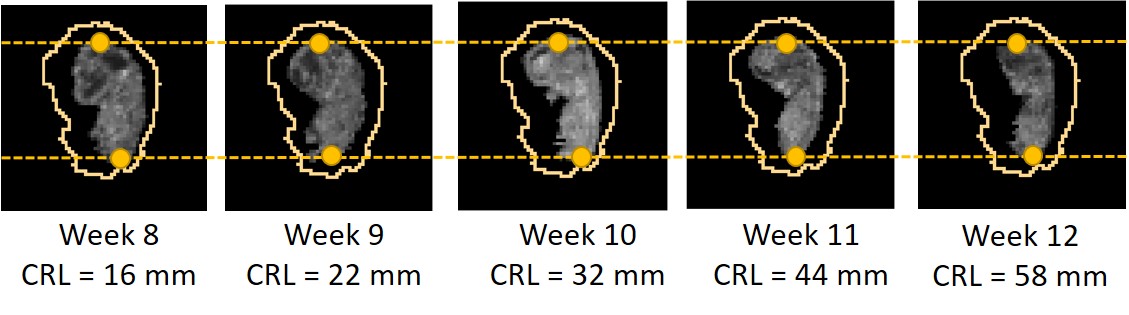}
    \caption{From left to right: example of an atlas at 8, 9, 10, 11 and 12 weeks GA, with their crown and rump landmarks. The outline shows the mask in which the similarity loss in \!\eqref{eq:ncc} is calculated. This mask is obtained by dilating the union of all atlas segmentations using a ball with radius 1.}
    \label{fig:aligned_atlas}
\end{figure}

\subsubsection{Learning to estimate the deformation $\phi$}
Firstly, note that our framework consists of two networks, one dedicated to learning an affine transformation and the other to learning a nonrigid deformation. In previous work we showed that separating these tasks was needed due to the wide variety in positions and orientation of the embryo \citep{Bastiaansen2020}. Therefore, the deformation $\phi$ consists of an affine transformation $\phi_{a}$ and a nonrigid deformation $\phi_{d}$, such that:
\begin{equation} \phi:=\phi_{a} \circ \phi_{d}. \end{equation}
The affine network, to estimate $\phi_a$, is trained in two stages, the first stage uses the crown and rump landmarks for supervision, while in the second stage the network is trained in an unsupervised way. The nonrigid network to estimate $\phi_d$ is trained in a fully unsupervised way. $\phi_{a}$ and $\phi_{d}$ are obtained by two convolutional neural networks (CNN), which model the functions $g_{\theta_{\text{affine}}}$: $(\phi_{a},I \circ \phi_a)=g_{\theta_{\text{affine}}}(I)$, with $\theta_{\text{affine}}$ the network parameters and $g_{\theta_{\text{nonrigid}}}$: $(\phi_{d},I_{\phi_{a}} \circ \phi_d)=g_{\theta_{\text{nonrigid}}}(I_{\phi_{a}},A)$, with $I_{\phi_{a}}=I \circ \phi_a$, and $\theta_{\text{nonrigid}}$ the network parameters. 

\subsubsection{The atlas $A$}
To deal with the rapid growth and variation in appearance of the embryo we propose to use a multi-atlas approach. We define $A_{i,j}$ to be an atlas, with $i=1,...,n_p$, where $n_p$ represents the number of subjects and $j \in W$, with $W$ the set of the weeks of pregnancy in which an ultrasound image is available. For example, atlas $A_{1,8}$ represent the ultrasound image of subject 1 acquired at 8 weeks GA.  Define $\mathcal{A}:=\{A_{i,j} |i = 1,...,n_p, \forall j \in W \}$ to be the set of all available atlases.

We manually rigidly aligned every atlas $A_{i,j}$ to the same standard orientation. Subsequently, all atlases, were scaled to the same size by matching the annotated crown and rump landmarks, as shown in Fig. \!\ref{fig:aligned_atlas}.
Aligning the crown and rump landmarks for every atlas has two benefits for training the neural network: 1) regardless of the GA the embryo always fills the same field of view, and 2) we only need to learn one affine transformation that aligns the images to every atlas.

We define the coordinates of the crown and rump landmarks in the standard orientation as $x^{\mathcal{A}}_c$ and $x^{\mathcal{A}}_r$. Finally, the segmentation $S_{A_{i,j}}$ was obtained manually.

\subsubsection{Atlas selection and fusion strategies}
For every image $I$ the $M$ atlases closest in GA are selected from the set of available atlases $\mathcal{A}$. Furthermore, from every subject maximally one atlas image is selected, therefore we have: $M\leq n_p$. We define three different atlas fusion strategies:
\begin{description}
    \item [\textbf{Single-subject strategy}] The framework uses atlases from a single subject $k$:
    \begin{align}
      \mathcal{A}_{k}^{s}&:=\{A_{i,j} | i=k, j \in W\}
  \end{align}
  with $M \leq n_p =1$.
  \item [\textbf{Multi-subject strategy}] The framework uses atlases from multiple subjects:
  \begin{equation}
    \mathcal{A}^m:=\{A_{i,j}|i=1,...,n_p, j \in W\}.
    \end{equation}
    with $M \leq n_p$
  \item [\textbf{Ensemble strategy}]The results of $n_p$ single-subject strategies are ensembled, hence\\ $M=n_p$. We call this strategy $\mathcal{A}^{e}$.
\end{description}
The set of available atlases and the $M$ selected atlases are the same during training and inference.

\subsubsection{Output of framework}
The framework outputs the segmentation of the embryo and the embryo in standard orientation. The segmentation $S_I$ is defined as the voxelwise majority vote over $S_i^{i,j}$. Where $S_i^{i,j}$ is defined as:
\begin{equation}\label{eq:final_seg}
    S_{I}^{i,j} : = S_{A_{i,j}} \circ (\phi^{-1}_{d_{i,j}} \circ \ \phi^{-1}_{a}).
\end{equation}
and $S_I$ as:
\begin{equation}
    S_I(x):=  \begin{cases}
    1      & \quad \text{if } \sum_{i=1}^M S_{I}^{i,j}(x)>\frac{M}{2}\\
  0  & \quad \text{else,}
  \end{cases}
\end{equation}
for all $M$ selected atlases.

The deformation to put the embryo in standard orientation is given by $\phi_a$ and the image in standard orientation is given by:
\begin{equation}
I_{\phi_a} := I \circ \phi_a
\end{equation}

\subsection{Affine registration network}
The input of the affine registration network is the image $I$. The affine registration network gives as output the estimated affine transformation $\phi_a$ and the corresponding affinely registered image $I \circ \phi_a$. Recall that we affinely registered every atlas to the same standard orientation. Hence, the affine transformation $\phi_a$ registers the image $I$ to all atlases in $\mathcal{A}$. Furthermore, this allowed us to directly compare the image similarity of $I \circ \phi_a$ and all selected atlas images in the loss function. Hence, as shown by the arrows in Fig. \!\ref{fig:network_archi}, in contrast to the nonrigid network, the atlas image $A_{i,j}$ is not given as input to the network.

\subsubsection{Network architecture}
The affine registration network consists of an encoder, followed by a global average pooling layer. The encoder consists of convolutional layers with a stride of 2, where the images are down-sampled. The numbers of filters $(f_1,f_2)$ are hyperparameters. The global average pooling layer gives as output one feature per feature map, which forces the network to encode position and orientation globally. The pooling layer is followed by four fully connected layers with 1000 neurons and ReLu activation. The output layer consists of the affine transformation $\phi_{a}$ and is defined as a $12$-dimensional vector containing the coefficients of part of the affine transformation matrix $T\in\mathbb{R}^{4 \times 4}$. 

\subsubsection{Loss function}
The loss function for the first training stage of the affine registration network is defined as:
\begin{align}\begin{split}\label{eq:loss1} \mathcal{L}\left(\mathcal{A}, I,\phi_a,x^\mathcal{A}, x^I\right)= \frac{1}{M}\sum_{i,j}\delta_{i,j}\mathcal{L}_{\text{sim}}\left(A_{i,j},I \circ \phi_a\right) \\
+ \lambda_{l} \mathcal{L}_{\text{landmark}}\left(\phi_a \left(x^\mathcal{A}\right),  x^I\right), \end{split}\end{align}
with $\delta_{i,j}$ defined as:
\begin{equation}
\delta_{i,j} =
  \begin{cases}
    1      & \quad \text{if } A_{i,j} \text{ selected}\\
  0  & \quad \text{else.}
  \end{cases}
\end{equation}
The first term of the loss function promotes similarity between the image $I$ after alignment and the atlas. The similarity loss is only calculated within a mask $\tilde\Omega$, since there are other objects in the 3D ultrasound images besides the embryo. Mask $\tilde\Omega$ is obtained by dilating the union of all segmentations $S_{A_{i,j}}\forall i,j$ using a ball with radius 1. $\mathcal{L}_{\text{sim}}(A,I\circ \phi_a)$ is chosen as the negative masked local (squared) normalized cross-correlation (NCC), which is defined as follows for the masked images $U,V$:
\begin{align}
\text{NCC}\left(U,V\right)=
\label{eq:ncc}\frac{1}{|\tilde\Omega|} \sum_{p \in \tilde\Omega}  \frac{\left ( \sum_{q} [U(q)-\bar{U}(p)][V(q)-\bar{V}(p)]\right)^2}{\left ( \sum_{q} [U(q)-\bar{U}(p)]^2\right)\left(\sum_{q} [V(q)-\bar{V}(p)]^2\right)}\end{align}
where $\bar{U}$ (and similarly for $\bar{V}$) denotes: $\bar{U}(p)=\frac{1}{j^3} \sum_{q} U(q)$, and where $q$ iterates over a $j^3$ volume around $p\in \Omega$ with $j=9$ as in \cite{Balakrishnan2018}. The second term in Eq. \!\eqref{eq:loss1} minimizes the mean Euclidean distance in voxels between the landmarks after registration and is defined as:
\begin{equation} \mathcal{L}_{\text{landmark}}\label{eq:tre}\left(\phi_a\left(x^\mathcal{A}\right), x^I\right)= \frac{1}{n_l} \sum_{i=1}^{n_l} \left\|x^I_i-\phi_a\left(x_i^\mathcal{A}\right)\right\|_2,\end{equation} 
with $n_l$ the number of annotated landmarks, with $n_l=2$ in our case.

During the second stage of training the weights $\theta_{\text{affine}}$ learned in the first stage are used as initialization of the network. The loss function for the second training stage of the affine registration network is defined as:
\begin{align}\begin{split}\label{eq:loss2} \mathcal{L}(\mathcal{A},I,\phi_a)=& \frac{1}{M}\sum_{i,j}\delta_{i,j}\mathcal{L}_{\text{sim}}\left(A_{i,j},I \circ \phi_a\right)\\ &+ \lambda_{s} \mathcal{L}_{\text{scaling}}\left(\phi_a\right), \end{split}\end{align}
When objects in the background are present, penalizing extreme zooming is beneficial, as was showed in \cite{Bastiaansen2020}. Hence, $\mathcal{L}_{\text{scaling}}$ is defined as:
\begin{equation} \mathcal{L}_{\text{scaling}}\left(\phi_a\right)=\sum_{i=1}^3 \log(s_i)^2, \end{equation}
following \cite{Ashburner1999}, with $s_i$ the scaling factors of $\phi_a$ obtained using the singular value decomposition.

\subsection{Nonrigid registration network}
The input of the nonrigid registration network is an affinely registered image $I_{\phi_a}:=I \circ \phi_a$ together with a selected atlas $A_{i,j}$. During training we provide as input to the network the image $I$ and one randomly chosen selected atlas $A_{i,j}$. During inference we give as input to the network all the possible pairs of image and selected atlas. The output of the network consists of $\phi_{d_{i,j}}$, along with the registered images $I_{\phi_a} \circ \phi_{d_{i,j}}$. 

To obtain the segmentation in Eq. \!\ref{eq:final_seg}, the deformation $\phi_{d_{i,j}}$ must be inverted. To ensure invertibility, we adopt diffeomorphic deformation for $\phi_{d_{i,j}}$. Following \cite{Dalca2019}, we use a stationary velocity field (SVF) representation, meaning that $\phi_{d_{i,j}}$ is obtained by integrating the velocity field $\nu$ \citep{Ashburner2007}. Now, $\phi_{d_i}^{-1}$ is obtained by integrating the velocity field using $-\nu$.

\subsubsection{Network architecture}
For the nonrigid registration network we used the architecture of Voxelmorph proposed by \cite{Balakrishnan2018}. Voxelmorph consists of an encoder,  with convolutional layers with a stride of 2, a decoder with convolutional layers with a stride of 1, skip-connections, and an up-sampling layer. This is followed by convolutional layers at full resolution, to refine the velocity field $\nu$. The numbers of filters $(f_3, f_4)$ are hyperparameters. The velocity field $\nu$ is integrated in the integration layer to give the dense displacement field. In both networks, all convolutional layers have a kernel size of 3 and have a LeakyReLU activation with parameter $0.2$.

\subsubsection{Loss function}
The loss function for the nonrigid registration network is:
\begin{equation}\label{eq:loss3} \mathcal{L}(A_{i,j},I_{\phi_a},\phi_{d_{i,j}})=\mathcal{L}_{\text{sim}}\left(A,I_{\phi_a}\circ \phi_{d_{i,j}} \right) +\lambda_{\text{d}} \mathcal{L}_{\text{diffusion}}\left(\phi_{d_{i,j}} \right), \end{equation}
where $I_{\phi_a(x)}:=I\circ\phi_a$ is the output of the affine network. $\mathcal{L}_{\text{sim}}$ is again defined as the NCC in Eq. \!\ref{eq:ncc}. $\phi_{d_{i,j}}$ is regularized by:
\begin{equation} \mathcal{L}_{\text{diffusion}}\left(\phi_{d_{i,j}}\right)= \frac{1}{|\tilde \Omega|}\sum_{p \in \tilde\Omega} \|\nabla d_{i,j}(p)\|^2,\end{equation}
This loss term penalizes local spatial variations in $\phi_{d_{i,j}}$ to promote smooth local deformations \citep{Balakrishnan2018}.

\subsection{Implementation details}
For training the ADAM optimizer was used with a learning rate of $10^{-4}$, as in \cite{Balakrishnan2018}. For the second training stage of the affine registration network, we lower the learning rate to $10^{-5}$, since the network has already largely converged in the first stage. We trained each part of the network for 300 epochs with a batch size of one. In the first training stage of the affine network data augmentation was applied. Each scan in the training set was either randomly flipped along one of the axes or rotated 90, 180 or 270 degrees. For the second stage and nonrigid network preliminary experiments showed that data augmentation did not improve the results. The framework is implemented using Keras \citep{chollet2015keras} with Tensorflow as backend \citep{Abadi2016}. The code is publicly available at:\\ \href{https://gitlab.com/radiology/prenatal-image-analysis/multi-atlas-seg-reg}{https://gitlab.com/radiology/prenatal-image-analysis/multi-atlas-seg-reg}

\section{Experiments}

\subsection{The Rotterdam Periconceptional Cohort}
The Rotterdam Periconceptional Cohort (Predict study) is a large hospital-based cohort study conducted at the Erasmus MC, University Medical Center Rotterdam, the Netherlands. This prospective cohort focuses on the relationships between periconceptional maternal and paternal health and embryonic and fetal growth and development \citep{Rousian2021, Steegers-Theunissen2016}. 3D ultrasound scans are acquired at multiple points in time during the first trimester. Ultrasound examinations were performed on a Voluson E8 or E10 (GE Healthcare, Austria) ultrasound machine. Furthermore, maternal and paternal food questionnaires are taken and other relevant data such as weight, height and bloodsamples are collected. 

\subsubsection{In- and exclusion criteria}
Women are eligible to participate in the Rotterdam Periconceptional Cohort if they were at least 18 years of age, with an ongoing singleton pregnancy less then 10 weeks of GA. Pregnancies were only included if the GA could be determined reliably and GA was calculated according to the first day of the last menstrual period (LMP), and in cases with an irregular menstrual cycle, unknown LMP or a discrepancy of more then seven days, GA was determined by the CRL measurements performed in the first trimester. In case of conception by means of in vitro fertilisation or intra-cytoplasmic sperm injection, the conception date was used to calculate the GA. We included pregnancies regardless of the mode of conception. The distribution of the pregnancy outcome was as follows: 83.3$\%$ no adverse pregnancy outcome, 2.5$\%$ congenital malformation, 0.7$\%$ postpartum death, $0.6\%$ intra-uterine fetal death, $0.2\%$ termination of pregnancy, $0.1\%$ stillbirth and $12.6\%$ unknown.

We included images that had sufficient quality to manually annotate the crown and rump landmarks. We included subjects with ultrasound images acquired between 8+0 and 12+6 weeks GA. This range was selected, since previous research by \cite{Rousian2013} showed that firstly the crown and rump landmarks could be reliably determined up until and including the 12th week of pregnancy, and secondly that other relevant measurements which are relevant for more in-depth insight in growth and development, such as head circumference, trans-cerebellar diameter, biparietal diameter and abdominal circumference can be measured reliable starting at a GA of 7+4. 

\subsubsection{Manual annotations and image pre-processing}
In the Rotterdam Periconceptional cohort the embryonic volume (EV) is semi-automatically measured using Virtual Reality (VR) techniques and a region-growing segmentation
algorithm that has been implemented specifically for this measurement \citep{rousian2009}. The embryonic volume was measured by various experienced researchers from the cohort over the years (2009-2018). Evaluation of this measurement showed that inter-observer variability between an experienced and inexperienced rater is 0.999 ($95\% CI: 0.997-0.999)$ and intra-observer variability was 0.999 ($95\% CI: 0.998-0.999)$ \citep{rousian2010}. We choose to use the available EV measurements for evaluation and refer to them as $EV_{gt}$.  Additionally, for 186 images corresponding segmentations were saved. We used these segmentation, referred to as $S_I^{gt}$, for evaluation as well.

The included images have isotropic voxel sizes varying between 0.06 mm and 0.75 mm. The resolution varied between 121 and 328 voxels in the first dimension, between 83 and 257 voxels in the second dimension and 109 and 307 voxels in the third dimension. To speed up training all images were down-sampled to $64 \times 64 \times 64$ voxels. To achieve this while keeping isotropic voxel sizing, all images were padded with zeros to a square shape and subsequently re-scaled to the right size.
\begin{figure}[b!]
    \centering
    \includegraphics[width=\textwidth]{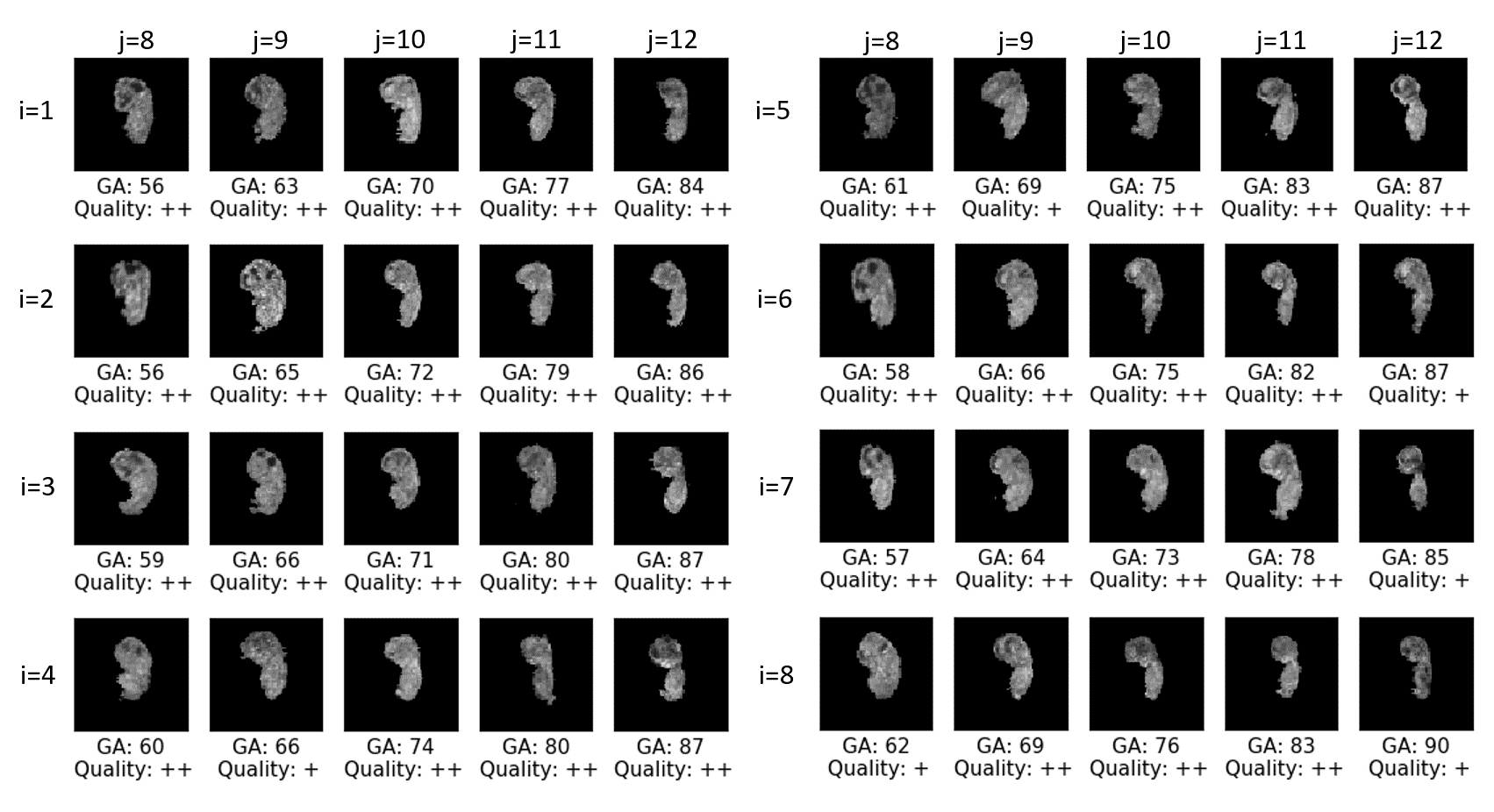}
    \caption{Overview of all atlases $A_{i,j}$ with $i=1,...,8$ the number of the subject and $J=8,9,10,11,12$ the week of pregnancy the ultrasound image was made. Below each image the GA is given along with the quality rating.  The quality is rating given during the measurements by the rater, distinguishing three categories: ++: excellent image quality, +: sufficient quality, -: low quality to perform measurements.}
    \label{fig:all_atlas}
\end{figure}
\subsubsection{Atlas, training, validation and test set}
We included data of 1282 subjects collected between December 2009 and August 2018. Eight subjects were used to create the atlases, the remaining 1274 subjects were divided as follows over the training set (874, 68\%), independent test set (200, 16\%) and validation set (200, 16\%) for hyperparameter optimization and fusion strategy comparison.

The eight subjects that formed the atlases were selected because they had weekly ultrasound images and had the highest overall image quality rating, this rating was assigned during the measurement in VR \citep{Rousian2013}. We choose to use eight atlases since adding more did not lead to improved performance in preliminary experiments. The choice to consider only subjects with weekly ultrasound images was made to ensure that we would have a good representation of every week GA among the atlases. In Fig. \!\ref{fig:all_atlas} a visualization of all atlases along with their GA in days and quality score can be found. We divided subjects over the train, validation and test set by following the rules below in the given order:
\begin{enumerate}
    \item all segmentations $S_I^{gt}$ must be divided equally over the validation and test set;
    \item all data of a subject should be in the same set;
    \item the majority of the EV measurements should correspond to ultrasound images in the validation or test set.
    \item within every week GA the division of unique subjects is approximately 60 \% - 20\% - 20\% over train, validation and test set.
\end{enumerate}   
In Tab. \!\ref{tab:charsplit} the characteristics of the final split can be found. 

\begin{table}[]
\caption{\label{tab:charsplit} Characteristics of the splitting of the data. Per gestational week the number of unique subjects (subj), ultrasound images (img), ultrasound images with embryonic volume measurement (EV) and ultrasound images with a ground truth segmentation (seg) is given. Note that per subject, in one or more weeks GA, one or more ultrasound images were available.}
\centering
\begin{tabular}{c|rrrr|rrrr|rrrr}
 & \multicolumn{4}{|c|}{Training} & \multicolumn{4}{|c|}{Validation}& \multicolumn{4}{|c}{Test}\\
 week & subj & img & EV & seg & subj & img & EV  & seg& subj & img & EV & seg\\
 \hline
 8 & 252 & 457 & 373 &  & 64 & 142 & 130 &  & 63 & 156 & 130 &  \\
 9 & 605 & 1333 & 1112 &  & 195 & 404 & 394 & 34 & 193 & 396 & 394 & 34 \\
 10 & 237 & 551 & 376 &  & 81 & 150 & 140 &  & 83 & 156 & 150 &  \\
 11 & 463 & 1190 & 865 & & 193 & 420 & 387 & 56 & 194 & 428 & 408 & 60 \\
 12 & 185 & 407 & 193 & & 68 & 122 &121 & & 67 & 113 & 113 & \\
 \hline
 all & 874 & 3938 & 2919 & & 200 & 1238 & 1172 & 90 & 200 & 1249 & 1205 & 94 \\
\end{tabular}
\end{table}
\subsection{Evaluation metrics}
The following four metrics were used to evaluate the results. Firstly, we report the relative error between the volume calculated from estimated segmentation $S_I$, referred to as $EV(I)$, and the ground truth embryonic volume $EV_{\text{gt}}(I)$:
\begin{equation} \label{eq:ev}\text{EV}_{\text{error}}(I):= \frac{\left |EV(I)-EV_{\text{gt}}(I)\right|}{EV_{\text{gt}}(I)}.\end{equation}

Secondly, we calculate the target registration error (TRE) in millimeters (mm) for the crown and rump landmarks as defined in the loss function in Eq. \!\ref{eq:tre}, to evaluate the accuracy of the affine alignment. The TRE is not calculated to evaluate the accuracy of the non-rigid alignment, since the TRE is only calculated in two landmarks, which is too limited.

Thirdly, we report the Dice similarity coefficient and the mean surface distance (MSD) in mm between the available manual segmentations $S_{I}^{\text{gt}}$ and the estimated segmentation $S_I$ of the image $I$ \citep{Dice}. 

To compare the results of different experiments, we performed the two-sided Wilcoxon signed-rank test with a significance level of 0.05, we report the p-values for the following three metrics as $p_{EV}$ and $p_{Dice}$ and $p_{MSD}$. We excluded the TRE from the statistical analysis, since the TRE only measures the alignment in two points, and therefore a lower TRE does not by definition means better alignment.

\subsection{Experiments}
We performed the following seven experiments. \\
\\
\textbf{Experiment 1: sensitivity metrics}\\
In literature there are no comparable studies available, to get insight in the performance of the presented framework we addressed the sensitivity to perturbations of the used evaluation metrics. We performed 4 operations on the 184 available ground truth segmentations: 1) dilation using a ball with radius of one voxel: $S_{gt} \oplus B_1$; 2) erosion using a ball with a radius of one voxel: $S_{gt} \ominus B_1$; 3) dilation using a ball with a radius of two voxels: $S_{gt} \oplus B_2$; 4) erosion using a ball with a radius of two voxels: $S_{gt} \ominus B_2$. We calculated for the four obtained segmentations the $\text{EV}_{\text{error}}$ and Dice score with respect to $S_{gt}$ and report the mean and standard deviation per operation.\\
\\
\textbf{Experiment 2: influence of hyperparameters and second affine training stage}\\
Using the single-subject strategy $\mathcal{A}_{1}^{s}$ we performed a hyperparameter search on the validation set for $\lambda_l \in \{0.01,1,100\}$ in Eq. \!\ref{eq:loss1}, $\lambda_s \in \{0, 0.005, 0.05, 0.5, 5\}$ in Eq. \!\ref{eq:loss2} and $\lambda_d \in \{1, 10, 100\}$ in Eq. \!\ref{eq:loss3} and the number of filters $(f_1, f_2)$ with $f_2=2 f_1$ and $f_1 \in \{16,32,64\}$ in the encoder of the affine network and $(f_3, f_4)$ with $f_4= 2 f_3$ and $f_3 \in\{1,2,4,8,16\}$ in the encoder and decoder of the nonrigid network. Furthermore, we took out training stage 2 of the affine network to evaluate its necessity. We call this strategy $\mathcal{A}_{1}^{s}$[no stage 2]. \\
\\
\textbf{Experiment 3: single-subject strategy}\\
We used the best hyperparameters from experiment 2 to train $\mathcal{A}_k^s$ for $k=2,3,...,8$ and compare the results on the validation set to the results for $\mathcal{A}^s_1$, to evaluate whether the choice of subject $k$ influences the results. Finally, we applied  $\mathcal{A}^s_k$ for best performing subject $k$ to the test set.\\
\\
\textbf{Experiment 4: comparison of the multi-subject and ensemble strategy}\\
To evaluate what is the best approach to combine data of multiple subjects we compared the results on the validation set of the multi-subject strategy $\mathcal{A}^{m}$ for $M=1$, $M=2$, $M=4$ and $M=8$ to the ensemble strategy $\mathcal{A}^e$. \\
\\
\textbf{Experiment 5: influence of atlas quality}\\
We investigated the influence of the quality of the ultrasound images on the results by excluding subject $i=8$, since this subject had the lowest image quality score (see Fig. \!\ref{fig:all_atlas}). We repeated experiment 4 without subject $i=8$ on the validation set and compared the results to the original results in experiment 4.\\
\\
\textbf{Experiment 6: comparison of the best models}\\
We compared the best strategy using data of multiple subjects to the best single-subject strategy on the test set. This was done to investigate whether utilizing data of multiple subjects improves the results. In addition, for the best performing model, we compared the results to the 3D nn-UNet \citep{isensee2021}. The nn-UNet was trained using: 1) the available ground truth segmentations of all atlases, and 2) additionally using the ground truth segmentations available from the validation set. This resulted in a training set of 40 and respectively 130 images. The nn-UNet was trained using the default parameters.\\
\\
\textbf{Experiment 7: analysis of best model}\\
In the last experiment we evaluated the best performing strategy for the proposed method in more detail. Recall from Tab. \!\ref{tab:charsplit} that in our dataset, for every subject at one or more weeks GA, one or more ultrasound images were acquired. These images vary in quality, and in orientation and position of the embryo. In clinical practice, only one of the acquired images is used for the measurement, hence the algorithm has to work for at least one of the available ultrasound images. Firstly, we visually scored the affine alignment for every ultrasound image as follows: 0: poor, 1: landmarks aligned, 2: acceptable, 3: excellent. For the further analysis we took per subject and week GA the best score. Secondly, we plotted for cases with scores 2 and 3 the obtained embryonic volume against the ground truth embryonic volume, to compare their distribution and calculate the correlation between them. Thirdly, we show a visualization of the resulting mid-sagittal after alignment and the obtained segmentation. Fourthly, to gain more insight into reasons why our framework might fail, we visually inspected all images in the test set {with a score of 0}.
\newpage
\section{Results}

\subsection{Experiment 1: sensitivity metrics}
We performed four operations ($S_{gt} \oplus B_1$, $S_{gt} \ominus B_1$, $S_{gt} \oplus B_2$, $S_{gt} \ominus B_2$) on the 184 available ground truth segmentations to investigate the sensitivity to perturbations. From Tab. \!\ref{tab:dileroexp} we conclude that an under- of overestimation of one voxel around the edge results in a Dice score between 0.81 and 0.84 on average and a $\text{EV}_{\text{error}}$ between 0.31 and 0.37 on average. An over- or underestimation of two voxels around the edge results in a Dice score between 0.61 and 0.71 and an $\text{EV}_{\text{error}}$ between 0.55 and 0.83. In the remaining experiments, these values serve as a reference for the residual errors.
\begin{table}[t!]
 \caption{Mean $\text{EV}_{\text{error}}$ and Dice score for the four operations. Standard deviation is given between brackets. The radius of the ball is given in voxels. }
    \begin{subtable}{0.45\textwidth}
    \caption{$\text{EV}_{\text{error}}$}
        \centering
        \begin{tabular}{c | l | l}
       Radius & Dilation & Erosion \\
        \hline 
        1 & 0.37 (0.06) & 0.31 (0.05) \\
        2 & 0.83 (0.17) & 0.55 (0.07) \\
       \end{tabular}
       \label{tab:evdilero}
    \end{subtable}
    \hfill
    \begin{subtable}{0.45\textwidth}
    \caption{Dice}
        \centering
        \begin{tabular}{c | l | l}
       Radius  & Dilation & Erosion \\
        \hline 
        1 & 0.84 (0.02) & 0.81 (0.03) \\
        2 & 0.71 (0.04) & 0.61 (0.07) \\
        \end{tabular}
        \label{tab:evdildice}
     \end{subtable}
     \label{tab:dileroexp}
\end{table}

\subsection{Experiment 2: influence of hyperparameters and second affine training stage}
\begin{figure}[t!]
    \centering
    \includegraphics[width=\textwidth]{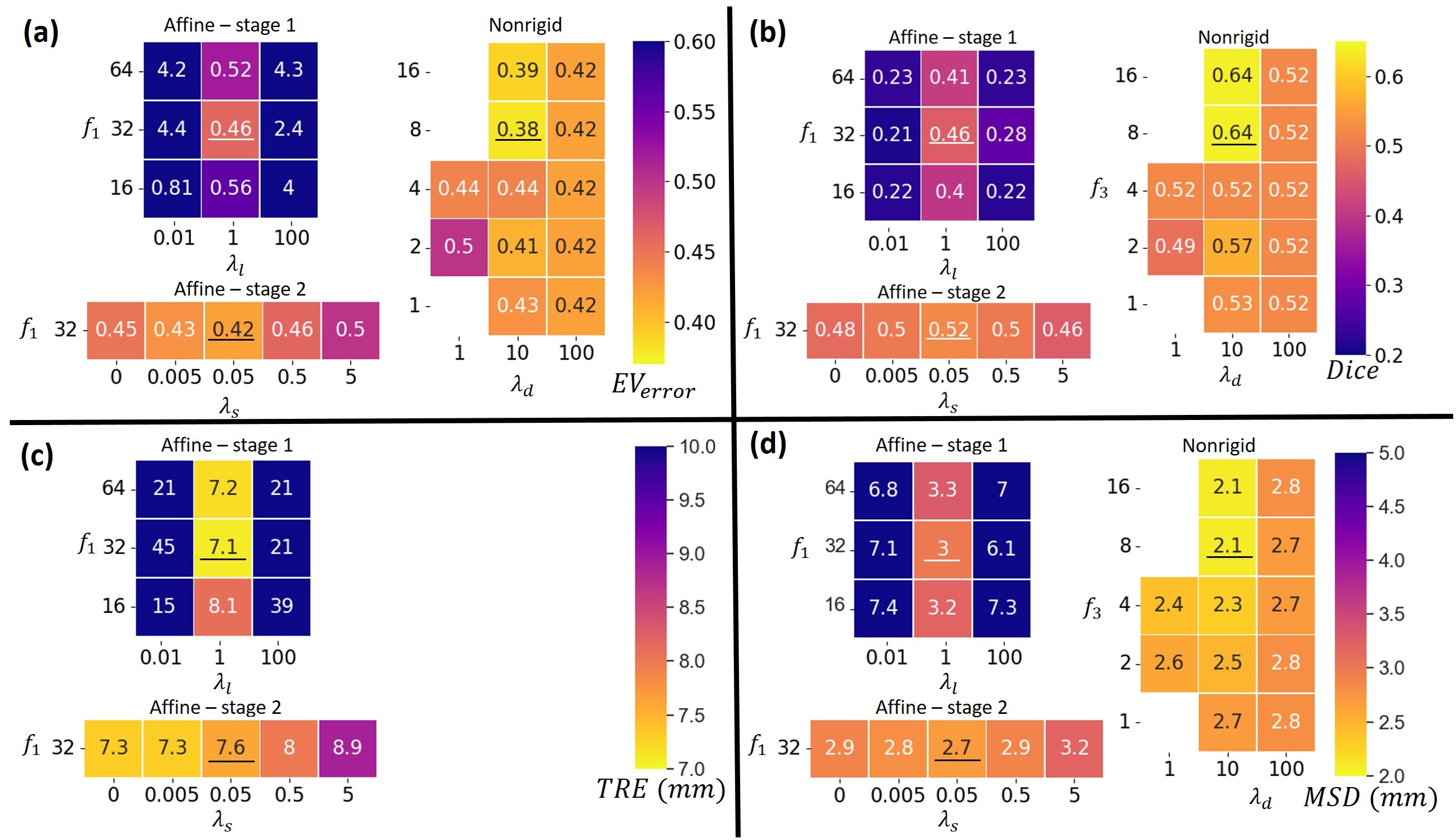}
    \caption{The mean $\text{EV}_{\text{error}}$ (a), mean Dice score (b), mean TRE (c) and mean MSD (d) for the validation set for different values of $\lambda_l$ in Eq. \!\eqref{eq:loss1}, $\lambda_s$ in Eq. \!\eqref{eq:loss2},  $\lambda_d$ in Eq. \!\eqref{eq:loss3} and the number of filters $(f_1,2f_1)$ in the affine network and $(f_3,2f_3)$ in the nonrigid network. The result for the chosen combination of hyperparameters is underlined. Hyperparameter combinations where the metrics became infinite are blank.}
    \label{fig:exp2_heatmap}
\end{figure}

In Fig. \!\ref{fig:exp2_heatmap} the mean $\text{EV}_{\text{error}}$, Dice score, TRE and MSD can be found for the validation set for different hyperparameters. For the first stage of the affine registration network, the best results were found for $\lambda_l=1$ and $(f_1,f_2)=(32,64)$. Setting $\lambda_l=0.01$ gave worse results, since the the supervision by landmarks was neglected. For $\lambda_l=100$ the results are worse, since in this case the image similarity term in Eq. \!\eqref{eq:loss1} is neglected in the optimization. 

Furthermore, Fig. \!\ref{fig:exp2_heatmap} shows that after applying the second training stage of the affine network in general the results improved. Only for $\lambda_s=5$ the results deteriorate, which is caused by penalizing scaling too much. The best results were found for $\lambda_s=0.05$. Note that the TRE increased a bit during the second training stage, which was to be expected, since the manually annotated landmarks have a limited precision and only measure alignment at two positions. Furthermore, we took $\lambda_s=0.05$ as optimal value, although for $\lambda_s=\{0,0.005\}$ the TRE is lower. This choice was made since the other metrics are optimal for $\lambda_s=0.05$, which indicates that the alignment in the two landmarks can be better with more scaling, but the overall quality deteriorates. 

Regarding the results of training the nonrigid registration network, Fig. \!\ref{fig:exp2_heatmap} shows that setting $\lambda_d=100$ results in low flexibility of the deformation field, which leads to no improvements of the results after nonrigid registration. Setting $\lambda_d=1$ resulted in more irregular deformation fields, which led to less improvement of the metrics compared to $\lambda_d=10$. The best results were found for $\lambda_d=10$ and $(f_3,f_4)=(8,16)$.

\begin{figure}[h!]
    \centering
    \includegraphics[width=\textwidth]{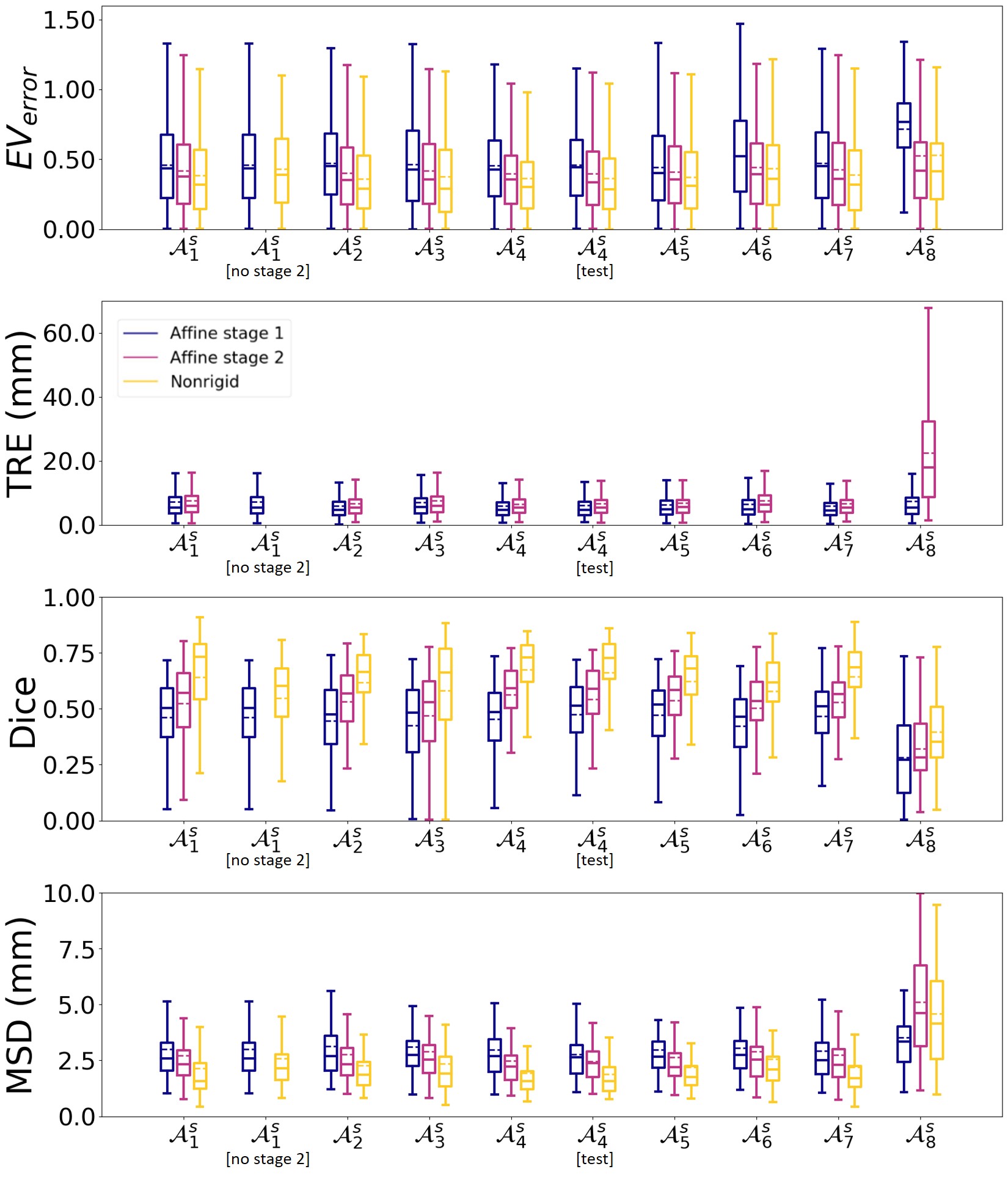}
    \caption{Results for all eight single-subject strategies evaluated over the validation set. $\mathcal{A}^{s}_1$[no stage 2] refers to the single-subject strategy where stage 2 of the affine network is omitted. The dotted line represents the mean. $\mathcal{A}^s_4$[test] refers to the results of evaluating $\mathcal{A}^s_4$ on the test set.}
    \label{fig:exp3a_boxplot}
\end{figure}

Finally, we compared using the results of stage 2 ($\mathcal{A}_1^s$) or the results of stage 1\\ ($\mathcal{A}_1^{s}$[no stage 2]) as input for the nonrigid registration network. Comparing the results for $\mathcal{A}_1^s$ and $\mathcal{A}^{s}_1$[no stage 2] in Fig. \!\ref{fig:exp3a_boxplot} shows that the results are better for all metrics for $\mathcal{A}_1^s$. The statistical test confirmed these observations ($p_{EV}<0.001$, $p_{Dice}<0.001$, $p_{MSD}<0.001$).
\subsection{Experiment 3: single-subject strategy}
Using the best hyperparameters determined in experiment 2, we trained the framework using the single-subject strategy for every subject $k=1,...,8$.Fig. \!\ref{fig:exp3a_boxplot} shows for the validation set that for $\mathcal{A}_k^s$, for $k=1,..,8$ the results improved between the stages of the affine network and after nonrigid registration in terms of the $\text{EV}_{\text{error}}$, Dice score and MSD. For the TRE we observed again in general a slight increase after the second training stage. Furthermore, a wide spread is observed in the results, which comes from cases where the affine registration fails and this can not be compensated for by the subsequent nonrigid registration network. 

In general our approach gives similar results independent of the choice of subject. However, there are differences in results between the different subjects, which are consistent with the variation in atlas image quality, since for example subject 8 had the least overall quality. The atlas image quality is given in Fig. \!\ref{fig:all_atlas}.

We applied the framework on the test set using the best hyperparameters found and the best single-subject strategy. Fig. \!\ref{fig:exp3a_boxplot} shows that $\mathcal{A}_4^s$ performs the best in terms of Dice score and TRE and comparable in terms of $\text{EV}_{\text{error}}$ and MSD. When applying this model to the test set we found that the results in Fig. \!\ref{fig:exp3a_boxplot} for $\mathcal{A}^s_4$[test] are comparable to those on the validation set ($\mathcal{A}^s_4$). 
\clearpage
\subsection{Experiment 4: comparison of the multi-subject and ensemble strategy}
\begin{figure}[t!]
    \centering
    \includegraphics[width=\textwidth]{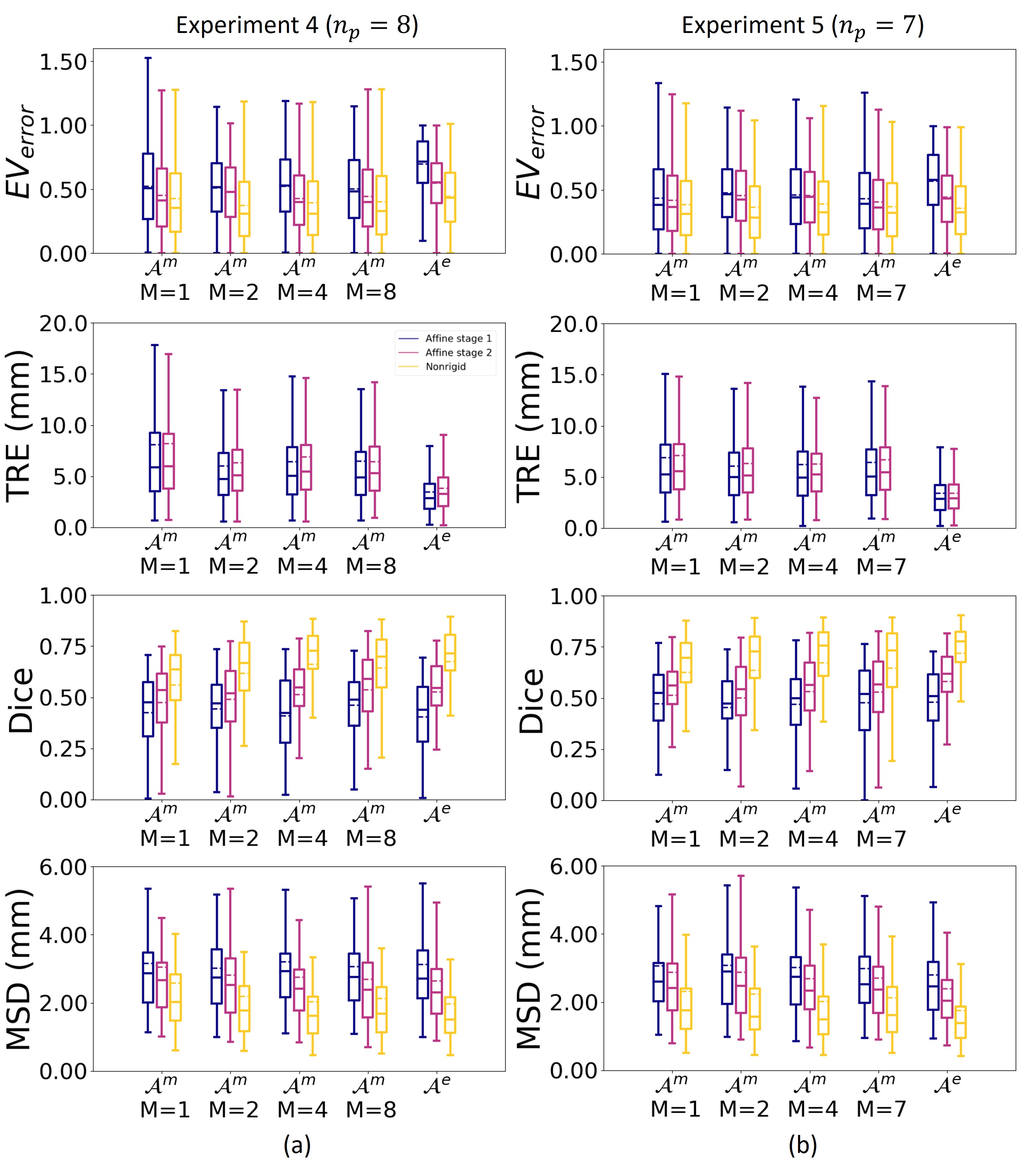}
    \caption{(a) Results for the validation set in experiment 4, where multi-subject strategy $\mathcal{A}_m$ for different values of $M$ is compared to the ensemble strategy $\mathcal{A}^e$. (b) Results for the validation set in experiment 5, where experiment 4 is repeated without the data of subject $i=8$ (which had the lowest image quality).} 
    \label{fig:exp34_boxplot}
\end{figure}

In experiment 4 we compared the multi-subject strategy $\mathcal{A}^m$, for different values of $M$, with the ensemble strategy $\mathcal{A}^e$. Fig. \!\ref{fig:exp34_boxplot}a shows that for the $\text{EV}_{\text{error}}$ using the multi-subject strategy $\mathcal{A}^m$ gave for all tested values of $M$ better results than taking the ensemble strategy $\mathcal{A}^e$. This is confirmed by the statistical tests, with the p-values in Tab. \!\ref{tab:test_exp2}. 

For the Dice score and MSD the ensemble strategy gave comparable results for the best multi-subject strategies with $M=\{4,8\}$.

For the $TRE$ we observe in Fig. \!\ref{fig:exp34_boxplot}a that the results for the ensemble strategy $\mathcal{A}^e$ are better than for the multi-atlas strategy $\mathcal{A}^m$. This was to be expected since the mean of $n_p=8$ affine transformations is taken, which reduces the error in the alignment. However, despite the better affine alignment, the segmentation did not improve. Hence, we conclude that the multi-subject strategy performs better.

Regarding the value of $M$ for the multi-subject strategy, we found that the results for $M=\{4,8\}$ significantly outperformed the results for $M=\{1,2\}$ for at least one metric. When comparing $M=4$ to $M=8$, we found that $M=4$ had a lower median value for the $\text{EV}_{\text{error}}$ (0.31 versus 0.33) and MSD (1.61 versus 1.69), and higher median value for the Dice score (0.73 versus 0.70). Hence we conclude that the multi-subject approach with $M=4$ is the best strategy to include data from multiple subject.

\begin{table}[b!]
    \caption{P-values of the two-sided Wilcoxon signed-rank test for the metrics on the validation set in experiment 4. Statistically significant results are boldface.}
    \label{tab:test_exp2}
    \centering
    \begin{tabular}{c|ccccc}
    $EV_{\text{error}}$ &  $\mathcal{A}^m, M=1$ & $\mathcal{A}^m, M=2$ &$\mathcal{A}^m, M=4$ & $\mathcal{A}^m, M=8$ & $\mathcal{A}^e$ \\
    \hline
    $\mathcal{A}^m, M=1$ &  & &  &  & \\
    $\mathcal{A}^m,M=2$ & $\mathbf{<0.001}$ & &  &  &\\
    $\mathcal{A}^m,M=4$ & $\mathbf{<0.001}$ & 0.780 & &  &  \\
   $\mathcal{A}^m,M=8$ & 0.064 & $\mathbf{0.005}$ & $0.51$ & &  \\
   $\mathcal{A}^e$ &$\mathbf{<0.001}$ & $\mathbf{<0.001}$ &$\mathbf{<0.001}$ & $\mathbf{<0.001}$ & \\
   \hline
    Dice&  $\mathcal{A}^m, M=1$ & $\mathcal{A}^m, M=2$ &$\mathcal{A}^m, M=4$ & $\mathcal{A}^m, M=8$ & $\mathcal{A}^e$ \\
     \hline 
     \\
   $\mathcal{A}^m, M=1$ &  & &  &  & \\
    $\mathcal{A}^m,M=2$ & $\mathbf{0.001}$ & &  &  &\\
    $\mathcal{A}^m,M=4$ & $\mathbf{<0.001}$ & $\mathbf{0.016}$ & &  &  \\
   $\mathcal{A}^m,M=8$ & $\mathbf{<0.001}$ & $0.147$ & $0.304$ & &  \\
   $\mathcal{A}^e$ &$\mathbf{<0.001}$ & $\mathbf{0.006}$ & $0.471$ & $0.052$ & \\
   \hline
   MSD&  $\mathcal{A}^m, M=1$ & $\mathcal{A}^m, M=2$ &$\mathcal{A}^m, M=4$ & $\mathcal{A}^m, M=8$ & $\mathcal{A}^e$ \\
     $\mathcal{A}^m, M=1$ &  &&  &  & \\
    $\mathcal{A}^m,M=2$ &  $\mathbf{<0.001}$& &  & & \\
    $\mathcal{A}^m,M=4$ & $\mathbf{<0.001}$ & $\mathbf{0.031}$  & &  &  \\
   $\mathcal{A}^m,M=8$ & $\mathbf{<0.001}$ & 0.812 & 0.071  & &  \\
   $\mathcal{A}^e$ &$\mathbf{<0.001}$ & $\mathbf{0.004}$ & 0.510  & $\mathbf{<0.017}$  & \\
    \end{tabular}
\end{table}
\newpage
\subsection{Experiment 5: influence of quality atlas images}
After excluding subject $i=8$ with the lowest image quality, experiment 4 was repeated for the other $n_p=7$ subjects. Fig. \!\ref{fig:exp34_boxplot}b shows that for multi-pregnancy strategy $\mathcal{A}^m$ with $M=1$ and the ensemble strategy $\mathcal{A}^e$ the results are significantly better when excluding the subject with lowest image quality ($\mathcal{A}^m$, $M=1$: $p_{EV}=0.001$, $p_{Dice}<0.001$, $p_{MSD}<0.001$, $\mathcal{A}^e$: $p_{EV}<0.001$, $p_{Dice}<0.001$, $p_{MSD}<0.001$). This can be explained by the fact that if fewer atlases are taken into account, atlases with worse quality will have more influence. 

When comparing the multi-subject strategy $\mathcal{A}^m$ with $M=7$ against the multi-subject strategy $\mathcal{A}^m$ with $M=8$, the $\text{EV}_{\text{error}}$, and was significantly better for $M=7$ ($p_{EV}=0.011$, $p_{Dice}=0.434$, $P_{MSD}=0.510$). For the multi-subject strategy with $M=2$ and $M=4$ the results were not significantly different for any metric ($M=2$: $p_{EV}=0.229$, $p_{Dice}=0.121$, $p_{MSD}=0.291$, $M=4$: $p_{EV}=0.404$, $p_{Dice}=0.445$, $p_{MSD}=0.741$).

Therefor, since the multi-subject strategy was not harmed by the inclusion of the 8th atlas subject with the worst quality, we conclude that the multi-subject strategy $A^m$, with $M=2$ or $M=4$ is more robust to choice of atlas.

\subsection{Experiment 6: comparison of the best models}
The results for the best single-subject strategy $\mathcal{A}^s_4$ and best multi-subject strategy $\mathcal{A}^m$ with $M=4$ were compared on the test set and no significant differences were found ($p_{EV}=0.75$, $p_{Dice}=0.71$, $p_{MSD}=0.86$). From the results we conclude that the multi-subject strategy with $M=4$ gives the best model, since this strategy omits choosing the right subjects as atlas. Furthermore, in experiment 5 we saw that the results are not affected by the presence of atlases with lower image quality. Finally, on the test set we find a median $\text{EV}_{\text{error}}$ of 0.29, Dice score of 0.72 and MSD of 1.58 mm.

\begin{figure}[b!]
    \centering
    \includegraphics[scale=0.25]{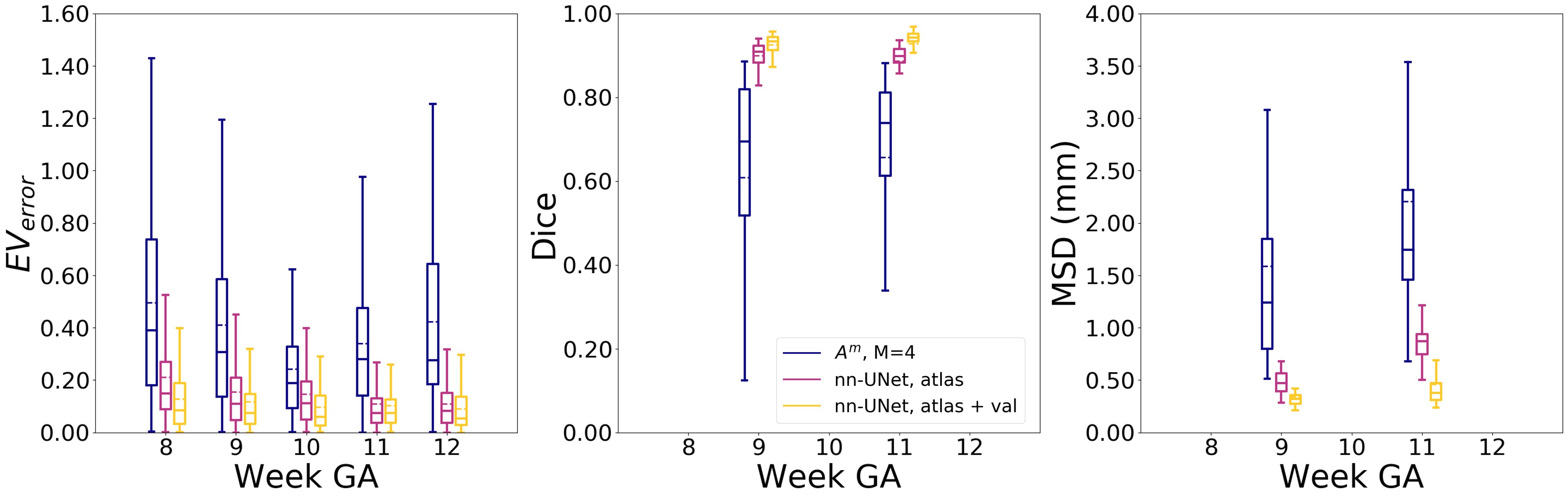}
    \caption{Result for comparison between the multi-subject strategy with $M=4$ and the full resolution 3D nn-UNet \cite{isensee2021} per week GA over the test set. The nn-UNet is trained on 1) the atlas subjects, 2) the atlas subjects plus images with a ground truth from the validation set.}
    \label{fig:nnunet}
\end{figure}

Fig. \!\ref{fig:nnunet} shows the results of the comparison with the 3D nn-UNet. The nn-UNet is trained on 1) the atlas subjects, 2) the atlas subjects plus images with a ground truth from the validation set. For every week GA, the nn-UNet outperformed our approach for segmentation.

\newpage

\subsection{Experiment 7: analysis of the best model}
We analyzed the results using the multi-subject strategy $\mathcal{A}^m$ with $M=4$ in more detail for the test set. Measuring the embryonic volume semi-automatically in VR takes between 5 and 10 minutes \citep{Rousian2013} per 3D image, the proposed framework takes on an Intel(R) Core(TM) i7-6850K CPU @ 3.60GHz 16 seconds to process one 3D image. Hence, we conclude that with appropriate hard- and software optimization our method can be used real-time in clinical practice.

\begin{figure}[b!]
    \centering
    \includegraphics[width=\textwidth]{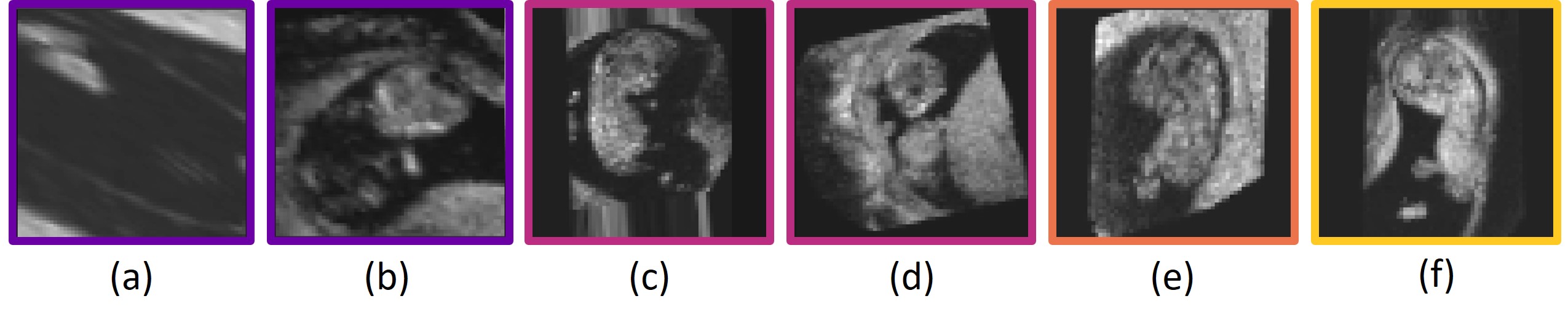}
    \caption{Examples of slices extracted at the mid-sagittal plane. (a,b) example for score 0: poor. (c,d) example for score 1: landmarks aligned. (e) example for score 2: acceptable. (f) example for score 3: excellent.}
    \label{fig:example_visual_score}
\end{figure}

In Tab. \!\ref{tab:visual} the results for the visual scoring are given. Overall, we found the following distribution: 0 (poor): 0.13 $\%$, 1 (landmarks aligned): 26 $\%$, 2 (acceptable): $39\%$, 3 (excellent): $21\%$. In Fig. \!\ref{fig:example_visual_score} examples for every score are given. 

\begin{figure}[b!]
    \centering
    \includegraphics[width=\textwidth]{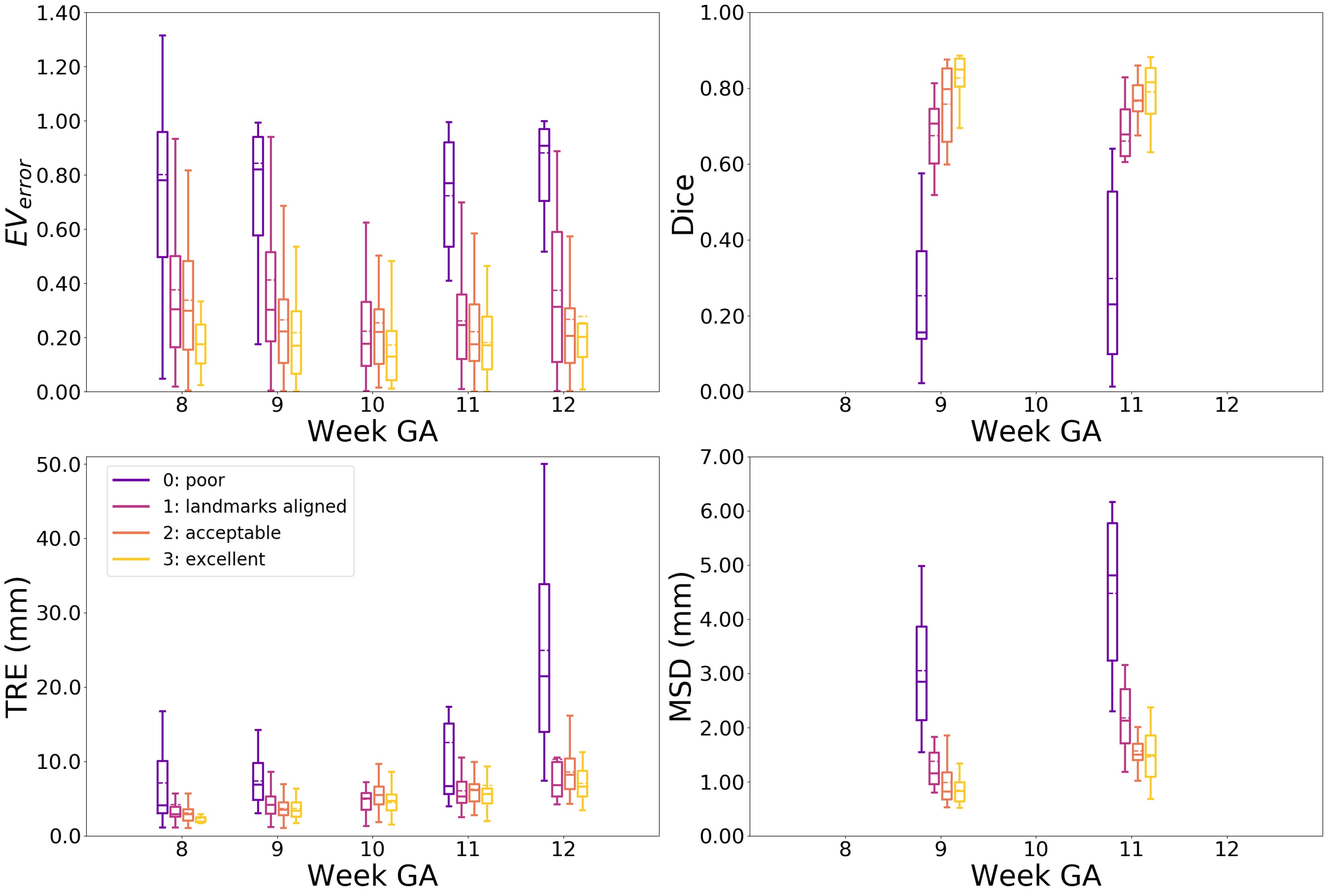}
    \caption{Results for visual alignment score for multi-subject strategy with $M=4$ per week GA for each metric.}
    \label{fig:visual_score}
\end{figure}

Next, we analyzed the metrics per score in Fig. \!\ref{fig:visual_score}. For each metric we observed that in general a better visual score corresponds to a lower $\text{EV}_{\text{error}}$, TRE, MSD, and a higher Dice score. Furthermore, we observe that the wide spread in the results for all metrics is many caused by failed affine alignment (score 0 and 1), and that in cases with successful alignment (score 2 and 3) the spread for the TRE, MSD and Dice score is smaller. For the $\text{EV}_{\text{error}}$ for scores 2 and 3 there is a wide spread, however this spread is comparable to the values found in experiment 1 and for the nn-UNet in experiment 6. 

Furthermore, we observed that all metrics are in the same range regardless of the week GA. Except for the $\text{EV}_{\text{error}}$ in week 8 and TRE in week 12. From the visual scoring in Tab. \ref{tab:visual} we found that week 8 has relative the most poorly aligned (score 0) cases, and this was reflected in the $\text{EV}_{\text{error}}$. For week 12 the TRE became the largest in case of poor alignment, which can be explained by the the larger size of the embryo in week 12.
\begin{table}[b!]
    \centering
    \begin{tabular}{c|c|c|c|c|c|c}
      & Week 8 & Week 9 & Week 10 & Week 11 & Week 12 & Total\\
      \hline
      0 & 22 (29$\%$) & 28 (15$\%$) & 0 (0$\%$) & 17 (9$\%$) & 7 (13$\%$) &74 (13$\%$) \\
      1 & 23 (31$\%$) & 46 (25$\%$) & 23 (37$\%$) & 45 (25$\%$) & 13 (23$\%$) &150 (26$\%$) \\
      2 & 23 (31$\%$) & 74 (40$\%$) & 27 (37$\%$) & 72 (39$\%$) & 28 (51$\%$)&224 (39$\%$) \\
      3 & 7 (9$\%$) & 36 (20$\%$) & 23 (32$\%$)& 49 (27$\%$) & 7 (13$\%$) &122 (21$\%$)\\
      \hline
      total & 75 & 184 & 73 & 183 & 55 & 570
    \end{tabular}
    \caption{Visual alignment score per week GA, with 0: poor, 1: landmarks aligned, 2: acceptable, 3: excellent. In the test set in there were 200 subjects with in one or more week GA, one or more ultrasound images taken.  We took per subject and week GA the image with the highest score, this resulted over 200 subject in 570 ultrasound images.}
    \label{tab:visual}
\end{table}

\begin{figure}[t!]
    \centering
    \includegraphics[width=\textwidth]{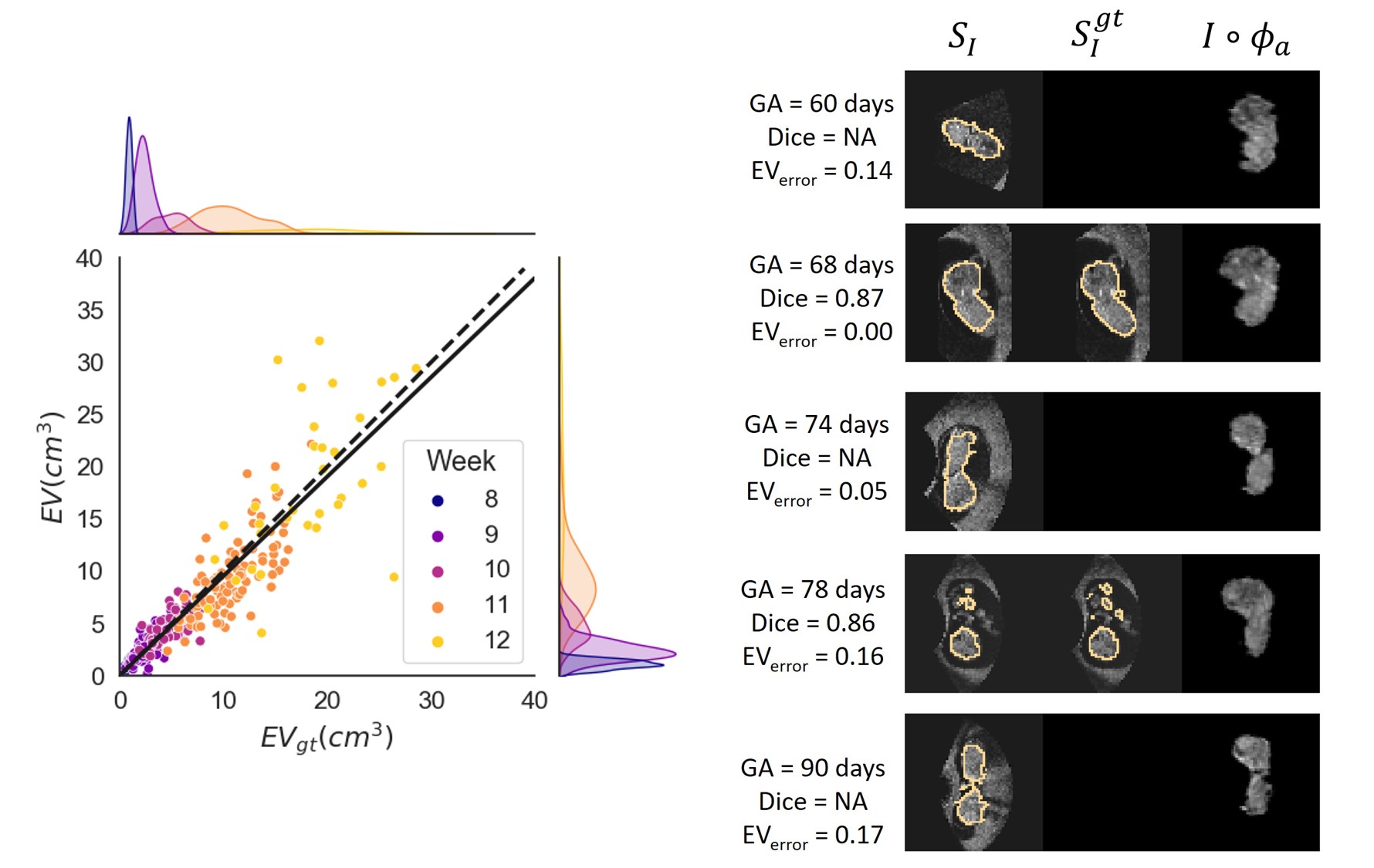}
    \caption{a) Joint plot for the $EV$ and $EV_{gt}$ in $\text{cm}^3$ for cases with a visual alignment score of 2 (acceptable) or 3 (excellent), for the multi-subject strategy $\mathcal{A}^m$ with $M=4$ for the test set. The dotted line represents $y=x$, the solid line is the result of linear regression: $y = 0.95x+0.04$. We found a correlation coefficient of 0.90. b) Results of multi-pregnancy strategy $\mathcal{A}^m$ with $M=4$ for the test set. For every week GA the mid-sagittal plane of an image from the test set is shown. The same slice is shown for every 3D volume.}
    \label{fig:M_4_week}
\end{figure}

Next, we plotted in Fig. \!\ref{fig:M_4_week}a, for cases with scores 2 and 3, the obtained embryonic volume against the ground truth. When performing linear regression on this data we found $y = 0.95x+0.04$ and a correlation of 0.90. This shows that despite the wide spread for the $\text{EV}_{\text{error}}$ in Fig.\! \ref{fig:visual_score}, we obtain accurate segmentations if the alignment is correct. Furthermore, fig. \!\ref{fig:M_4_week}b shows for every gestational week the mid-sagittal plane of an image from the test set. GA, $\text{EV}_{\text{error}}$ and Dice score are given per image.
\newpage

\begin{figure}[t!]
    \centering
    \includegraphics[width=\textwidth]{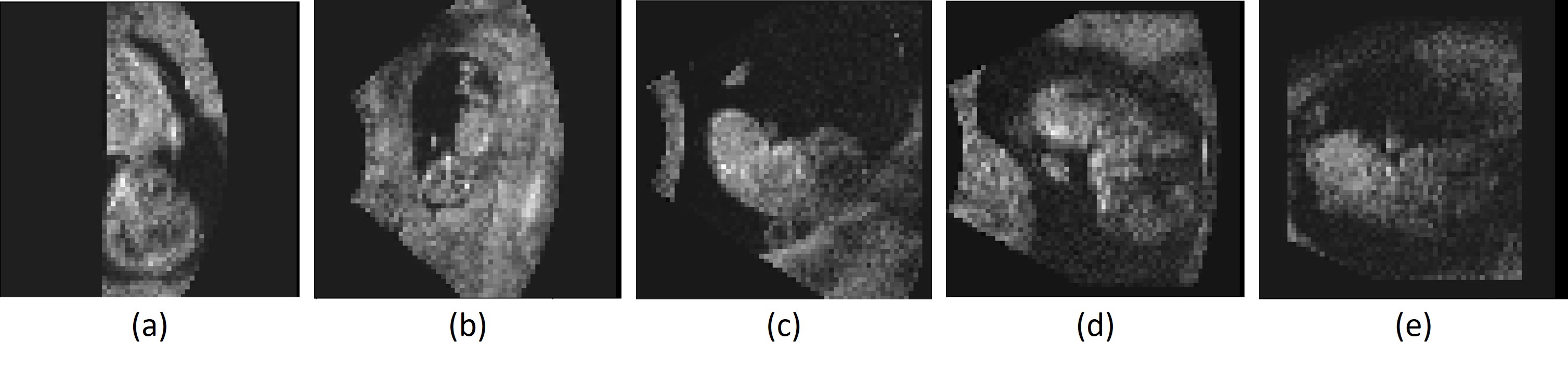}
    \caption{Examples of the original image for which the the affine registration network failed. (a) example of an image where the embryo lies in the image border. (b,c) examples of images where the embryo lies against the uterine wall and there is no clear distinction between embryo and uterine wall. (d,e) examples of images with low quality.}
    \label{fig:failed_affine}
\end{figure}

Finally, we visually inspected all images in the test set with a visual alignment score of 0. We observed that in $8\%$ of the images the embryo was lying against the border of the image, in $47\%$ of the images the embryo was lying against the uterine wall, in $34\%$ the image quality was very low, and in $11\%$ of the images there was no apparent visual cue why the affine registration failed. Fig. \!\ref{fig:failed_affine} shows some examples.

\clearpage

\section{Discussion and conclusion}
Monitoring first trimester embryonic growth and development using ultrasound is of crucial importance for the current and future health of the fetus. A first key step in automating this monitoring, is achieving automatic segmentation and spatial alignment of the embryo. Automation of those tasks will result in less investigation time and more consistent results \citep{Carneiro2008,blaas2006}. Here, we present the first framework to perform both tasks simultaneously for first trimester 3D ultrasound images. We have developed an multi-atlas segmentation framework and showed that we could segment and spatially align the embryo captured between 8+0 and 12+6 weeks GA. 

We evaluated three atlas fusion strategies for deep learning based multi-atlas segmentation, namely: 1) single-subject strategy: training the framework using atlas images from a single subject, 2) multi-subject strategy: training the framework with data of all available atlas images and 3) ensemble strategy: ensembling of the frameworks trained per subject. The multi-subject strategy significantly outperformed the ensembling strategy and the best results on were found when taking the $M=4$ atlases closest in gestational age in account. The best single-subject strategy and the best multi-subject strategy did not significantly outperform each other. However, the multi-subject strategy circumvents the need for selecting the most suitable subject to create atlases and is therefore a more robust method.

We took a deep learning approach for image registration. Our framework consist of two networks, the first one dedicated to learning the affine transformation and the second one to learning a nonrigid registration. The affine network is trained in two stages: the first stage is trained supervised using the crown and rump landmarks and subsequently in the second stage the results is refined in an unsupervised manner. In our experiments using the single-subject strategy we showed that this second stage significantly improves, as expected, the final result over skipping this stage. 

Furthermore, in the single-subject experiment we found that, regardless of the subject used, the result improves after each training step, but atlases with a higher image quality gave better results. From this we conclude that the choice of atlas subjects matters for the results.

When using multiple subjects, this choice is circumvented. Here, we found that the multi-subject strategy outperformed the ensemble strategy significantly in terms of the $\text{EV}_{\text{error}}$. For the Dice score and MSD both strategies performed similarly. Note that in terms of the TRE, the ensemble strategy gave the best results. However, this improvement in alignment is not reflected in the significantly inferior segmentation results. Furthermore, when excluding the subject that gave the worst results as single-subject strategy, we found that for the ensemble strategy the result, for all metrics, significantly improved. For the multi-subject strategy this was not the case. Hence, we conclude that taking a multi-subject strategy is more robust.

The robustness for choice of atlas subject of the multi-subject strategy can be explained by the fact that one network is trained using all available subjects. This has two benefits: firstly, the network can use the variation among atlases to model the variation present in the data, and secondly similarity to the atlas after registration is considered for all subjects, which reduces the influence of individual quality of atlas images. For the ensemble strategy the network is only trained on one subject, which does not allow for modelling variation among atlases.

In the experiments we evaluated for the multi-subject strategy how many $M$ atlases should be used for registration. The best results were found for $M=4$, this is explained by the fact that taking $M<4$ may not present enough variation to model the appearance of every image $I$. Surprisingly, taking $M>4$ did not improve the results. This could be explained by the fact that the subjects chosen as atlas were selected based on quality and on covering almost every day between 8 and 12 weeks gestation, see Fig. \!\ref{fig:all_atlas}. When taking $M>4$, this leads to a comparison with atlases covering a wider age range. This can be problematic for alignment, since these atlases may show different stages of development.

Finally, we analyzed in-depth the best model. Firstly, we compared the proposed method to the 3D nn-UNet \citep{isensee2021}, and we found that nn-UNet outperformed our method. The main difference between nn-UNet and our method is that we predict the deformation field to registers the image to a standard orientation, and from there derive the segmentation. The main limitation of our work is that when the affine registration fails, the nonrigid network cannot correct for this. This resulted in a wide spread in the results that was propagated from the affine to the nonrigid network for all metrics, and all atlas fusion strategies. To overcome this limitation, a good direction for further research to improve the alignment results are to either add additional landmarks, add the embryonic volume error to the loss, or incorporate the segmentations obtained by nn-UNet to improve the alignment results. These segmentations can be incorporated either a priori by segmenting the embryo before alignment, or can be added to the loss as supervision. Although it is shown in literature that including supervision using segmentations improve the results for nonrigid registration in general \citep{Balakrishnan2018}, due to wide variation in position and orientation of the embryo, the affine registration might remain challenging to find despite having the segmentation available.

We visually scored the alignment quality, and we found that overall for 39$\%$ of the test cases the result was acceptable and for 21$\%$ excellent. Furthermore, for the acceptable to excellent cases we found a correlation between the obtained EV and ground truth EV of $0.90$, indicating a strong correlation. Hence we conclude that when the alignment of the embryo is successful, we can accurately obtain the segmentation. Furthermore, we observed the cases where the alignment failed, that there was a clear visual explanation for the failing of the registration: either the embryo lying on the image border, against the uterine wall or the image had low quality.

The main strength of our work is that besides segmentation of the embryonic volume, we spatially align the embryo to a predefined standard orientation. Within this standard space any standard plane with there corresponding biometric measurement can easily be extracted. In future work, we aim to evaluate the proposed method for other measurements done within the Rotterdam Periconceptional cohort, for example the head volume, biparietal diameter and occipital-frontal diameter \citep{Rousian2018}.

A point for discussion is the usage of an affine transform for alignment to the standard space, where a rigid or similarity transform could be more appropriate to define a standard space. We initially choose to use an affine transformation to have the best input for the subsequent nonrigid deformation network, since we observed in previous work that the nonrigid network was not able to capture this \citep{Bastiaansen2020}.

Regarding the pre-processing of the data, we chose to down-sample all data to 64x64x64 voxels to speed up the training. However, this may have resulted in loss of information that might have influenced the results. We did not evaluate if using images at a higher resolution gave better results. 

In our experiments we chose to train separate networks per atlas subject. However, another interesting approach could be to train separate networks per week GA. We did not investigate this, since our aim was to develop one framework for the whole GA range. Furthermore, dividing the images per week GA might negatively influence the results of the images with a GA of for example 8+6, where the atlases for week 9 might be more similar then the ones for week 8.

The extension to non-singleton pregnancies is an interesting topic for further research, which was not considered here. For our framework, this implies that it should be able to deal with the presence of another (partial) embryo, or the possibility to align them both independently. We did not include this case in the current work, because non-singleton pregnancies are excluded from the Rotterdam Periconceptional cohort.

The proposed framework can also be used to support growth and development modelling, both for normal development and in the case of adverse outcomes, such as miscarriage, intra-uterine fetal death or postpartum death. Here, we did not further investigate this since only 52 (4\%) of the pregnancies had an adverse outcome. However, we used the proposed framework to segment and spatially align the embryonic brain for the development of a spatio-temporal model that can be used to explore correlations between maternal periconceptional health and brain growth and development \citep{bastiaansen2022}.

Finally, another interesting topic for further research is applying our framework to other problems. We created a flexible framework that easily can be adapted to work with or without landmarks and with or without multiple atlas images. Furthermore, these atlases could be longitudinal, like presented here, but the framework can also be applied to cross-sectional atlases. The atlas selection is currently based on GA, but could be based on any other relevant meta-data.

We conclude that the presented multi-atlas segmentation and alignment framework enables the first key steps towards automatic analysis of first trimester 3D ultrasound. Automated segmentation and spatial alignment pave the way for automated biometry and volumetry, standard plane detection, abnormality detection and spatio-temporal modelling of growth and development. This will allow for more accurate and less time-consuming early monitoring of growth and development in this crucial period of life.



%

\ethics{This study was approved by the local Medical Ethical and Institutional Review Board of the Erasmus MC, University Medical Center, Rotterdam, The Netherlands. Prior to participation, all participants provided written informed consent.}

\newpage
\coi{Wiro Niessen is founder, shareholder and scientific lead of Quantib BV. The authors declare no other conflicts of interest.}

\bibliography{refs_melba}

\begin{thebibliography}{44}
\providecommand{\natexlab}[1]{#1}
\providecommand{\url}[1]{\texttt{#1}}
\expandafter\ifx\csname urlstyle\endcsname\relax
  \providecommand{\doi}[1]{doi: #1}\else
  \providecommand{\doi}{doi: \begingroup \urlstyle{rm}\Url}\fi

\bibitem[Abadi et~al.(2016)Abadi, Agarwal, Barham, Brevdo, Chen, Citro,
  Corrado, Davis, Dean, Devin, Ghemawat, Goodfellow, Harp, Irving, Isard, Jia,
  Jozefowicz, Kaiser, Kudlur, Levenberg, Mane, Monga, Moore, Murray, Olah,
  Schuster, Shlens, Steiner, Sutskever, Talwar, Tucker, Vanhoucke, Vasudevan,
  Viegas, Vinyals, Warden, Wattenberg, Wicke, Yu, and Zheng]{Abadi2016}
M.~Abadi, A.~Agarwal, P.~Barham, E.~Brevdo, Z.~Chen, C.~Citro, G.~S. Corrado,
  A.~Davis, J.~Dean, M.~Devin, S.~Ghemawat, I.~Goodfellow, A.~Harp, G.~Irving,
  M.~Isard, Y.~Jia, R.~Jozefowicz, L.~Kaiser, M.~Kudlur, J.~Levenberg, D.~Mane,
  R.~Monga, S.~Moore, D.~Murray, C.~Olah, M.~Schuster, J.~Shlens, B.~Steiner,
  I.~Sutskever, K.~Talwar, P.~Tucker, V.~Vanhoucke, V.~Vasudevan, F.~Viegas,
  O.~Vinyals, P.~Warden, M.~Wattenberg, M.~Wicke, Y.~Yu, and X.~Zheng.
\newblock {TensorFlow: Large-Scale Machine Learning on Heterogeneous
  Distributed Systems}.
\newblock 2016.
\newblock URL \url{http://arxiv.org/abs/1603.04467}.

\bibitem[Al-Bander et~al.(2019)Al-Bander, Alzahrani, Alzahrani, Williams, and
  Zheng]{al2019}
B.~Al-Bander, T.~Alzahrani, S.~Alzahrani, B.M. Williams, and Y.~Zheng.
\newblock Improving fetal head contour detection by object localisation with
  deep learning.
\newblock In \emph{Annual Conference on Medical Image Understanding and
  Analysis}, pages 142--150, 2019.

\bibitem[Ashburner(2007)]{Ashburner2007}
J.~Ashburner.
\newblock {A fast diffeomorphic image registration algorithm}.
\newblock \emph{NeuroImage}, 38:\penalty0 95--113, 2007.

\bibitem[Ashburner et~al.(1999)Ashburner, Andersson, and
  Fristen]{Ashburner1999}
J.~Ashburner, J.~L.R. Andersson, and K.~J. Fristen.
\newblock {Image registration using a symmetric prior- in three-dimensions}.
\newblock \emph{NeuroImage}, 9:\penalty0 212--225, 1999.

\bibitem[Balakrishnan et~al.(2019)Balakrishnan, Zhao, Sabuncu, Guttag, and
  Dalca]{Balakrishnan2018}
G.~Balakrishnan, A.~Zhao, M.~R. Sabuncu, J.~Guttag, and A.~V. Dalca.
\newblock {Voxelmorph: a learning framework for deformable medical image
  registration}.
\newblock \emph{IEEE TMI}, 38:\penalty0 1788--1800, 2019.

\bibitem[Bastiaansen et~al.(2020{\natexlab{a}})Bastiaansen, Rousian,
  Steegers-Theunissen, Niessen, Koning, and Klein]{Bastiaansen2020}
W.A.P. Bastiaansen, M.~Rousian, R.P.M. Steegers-Theunissen, W.J. Niessen,
  A.~Koning, and S.~Klein.
\newblock Towards segmentation and spatial alignment of the human embryonic
  brain using deep learning for atlas-based registration.
\newblock In \emph{Biomedical Image Registration}, pages 34--43,
  2020{\natexlab{a}}.

\bibitem[Bastiaansen et~al.(2020{\natexlab{b}})Bastiaansen, Rousian,
  Steegers-Theunissen, Niessen, Koning, and Klein]{Bastiaansen2020a}
W.A.P. Bastiaansen, M.~Rousian, R.P.M. Steegers-Theunissen, W.J. Niessen, Anton
  Koning, and Stefan Klein.
\newblock Atlas-based segmentation of the human embryo using deep learning with
  minimal supervision.
\newblock In \emph{Medical Ultrasound, and Preterm, Perinatal and Paediatric
  Image Analysis}, pages 211--221. Springer, 2020{\natexlab{b}}.

\bibitem[Bastiaansen et~al.(2022)Bastiaansen, Rousian, Steegers-Theunissen,
  Niessen, A., and Klein]{bastiaansen2022}
W.A.P. Bastiaansen, M.~Rousian, R.P.M. Steegers-Theunissen, W.~Niessen, Koning
  A., and S.~Klein.
\newblock Towards a 4d spatio-temporal atlas of the embryonic and fetal brain
  using a deep learning approach for groupwise image registration.
\newblock In \emph{Submitted to 10th Internatioal Workshop on Biomedical Image
  Registration}, 2022.
\newblock URL \url{https://openreview.net/forum?id=HVCDHNRryh-}.

\bibitem[Blaas et~al.(2006)Blaas, Taipale, Torp, and Eik-Nes]{blaas2006}
H.G.K. Blaas, P.~Taipale, H.~Torp, and S.H. Eik-Nes.
\newblock Three-dimensional ultrasound volume calculations of human embryos and
  young fetuses: a study on the volumetry of compound structures and its
  reproducibility.
\newblock \emph{Ultrasound in Obstetrics and Gynecology: The Official Journal
  of the International Society of Ultrasound in Obstetrics and Gynecology},
  27\penalty0 (6):\penalty0 640--646, 2006.

\bibitem[Boveiri et~al.(2020)Boveiri, Khayami, Javidan, and
  Mehdizadeh]{Boveiri2020}
H.R. Boveiri, R.~Khayami, R.~Javidan, and A.~Mehdizadeh.
\newblock Medical image registration using deep neural networks: A
  comprehensive review.
\newblock \emph{Computers \& Electrical Engineering}, 87:\penalty0 106767,
  2020.

\bibitem[Carneiro et~al.(2008)Carneiro, Georgescu, Good, and
  Comaniciu]{Carneiro2008}
G.~Carneiro, B.~Georgescu, S.~Good, and D.~Comaniciu.
\newblock Detection and measurement of fetal anatomies from ultrasound images
  using a constrained probabilistic boosting tree.
\newblock \emph{IEEE Trans Med Imaging}, 27\penalty0 (9):\penalty0 1342--1355,
  2008.

\bibitem[Chen et~al.(2012)Chen, Tsai, Huang, Shih, Wang, Chang, and
  Sun]{Chen2012}
H.~C. Chen, P.~Y. Tsai, H.~H. Huang, H.~H. Shih, Y.~Y. Wang, C.~H. Chang, and
  Y.~N. Sun.
\newblock {Registration-Based Segmentation of Three-Dimensional Ultrasound
  Images for Quantitative Measurement of Fetal Craniofacial Structure}.
\newblock \emph{Ultrasound Med Biol}, 38:\penalty0 811--823, 2012.

\bibitem[Chollet and et~al.(2015)]{chollet2015keras}
F.~Chollet and et~al.
\newblock Keras, 2015.
\newblock URL \url{https://github.com/fchollet/keras}.

\bibitem[Dalca et~al.(2019)Dalca, Balakrishnan, Guttag, and Sabuncu]{Dalca2019}
A.~V. Dalca, G.~Balakrishnan, J.~Guttag, and M~.R. Sabuncu.
\newblock Unsupervised learning of probabilistic diffeomorphic registration for
  images and surfaces.
\newblock \emph{Med Image Anal}, 57:\penalty0 226--236, 2019.

\bibitem[de~Vos et~al.(2019)de~Vos, Berendsen, Viergever, Sokooti, Staring, and
  I{\v{s}}gum]{DeVos2019}
B.~D. de~Vos, F.~F. Berendsen, M.~A. Viergever, H.~Sokooti, M.~Staring, and
  I.~I{\v{s}}gum.
\newblock {A deep learning framework for unsupervised affine and deformable
  image registration}.
\newblock \emph{Med Image Anal}, 52:\penalty0 128--143, 2019.

\bibitem[Dice and Lee(1945)]{Dice}
F.~Dice and R.~Lee.
\newblock Measures of the amount of ecologic association between species.
\newblock \emph{Ecology}, 26\penalty0 (3):\penalty0 297--302, 1945.

\bibitem[Ding et~al.(2019)Ding, Han, and Niethammer]{Ding2019}
Z.~Ding, X.~Han, and M.~Niethammer.
\newblock Votenet: A deep learning label fusion method for multi-atlas
  segmentation.
\newblock In \emph{Medical Image Computing and Computer Assisted Intervention
  -- MICCAI 2019}, pages 202--210, 2019.

\bibitem[Droste et~al.(2020)Droste, Drukker, Papageorghiou, and
  Noble]{droste2020}
R.~Droste, L.~Drukker, A.T. Papageorghiou, and J~A. Noble.
\newblock Automatic probe movement guidance for freehand obstetric ultrasound.
\newblock In \emph{International Conference on Medical Image Computing and
  Computer-Assisted Intervention}, pages 583--592, 2020.

\bibitem[Fang et~al.(2019)Fang, Zhang, Nie, Cao, Rekik, Lee, He, and
  Shen]{Fang2019}
L.~Fang, L.~Zhang, D.~Nie, X.~Cao, I.~Rekik, S.~W. Lee, H.~He, and D.~Shen.
\newblock {Automatic brain labeling via multi-atlas guided fully convolutional
  networks}.
\newblock \emph{Med Image Anal}, 51:\penalty0 157--168, 2019.

\bibitem[Gutiérrez-Becker et~al.(2013)Gutiérrez-Becker, Arámbula, Guzmán,
  Benavides-Serralde, Camargo-Marín, and Medina]{Gutierrez2013}
B.~Gutiérrez-Becker, C.F. Arámbula, H.M.E. Guzmán, J.A. Benavides-Serralde,
  L.~Camargo-Marín, and B.V. Medina.
\newblock Automatic segmentation of the fetal cerebellum on ultrasound volumes,
  using a 3d statistical shape model.
\newblock \emph{Med Biol Eng Comput}, 51\penalty0 (9):\penalty0 1021--30, 2013.

\bibitem[Iglesias and Sabuncu(2015)]{Iglesias2015}
J.~E. Iglesias and M.~R. Sabuncu.
\newblock {Multi-atlas segmentation of biomedical images: A survey}.
\newblock \emph{Med Image Anal}, 24:\penalty0 205--219, 2015.

\bibitem[Isensee et~al.(2021)Isensee, Jaeger, Kohl, Petersen, and
  Maier-Hein]{isensee2021}
F.~Isensee, P.F Jaeger, S.A.A. Kohl, J.~Petersen, and K.H. Maier-Hein.
\newblock nnu-net: a self-configuring method for deep learning-based biomedical
  image segmentation.
\newblock \emph{Nature methods}, 18\penalty0 (2):\penalty0 203--211, 2021.

\bibitem[Kuklisova-Murgasova et~al.(2013)Kuklisova-Murgasova, Cifor,
  Napolitano, Papageorghiou, Quaghebeur, Rutherford, Hajnal, Noble, and
  Schnabel]{Kuklisova-Murgasova2013}
M.~Kuklisova-Murgasova, A.~Cifor, R.~Napolitano, A.~Papageorghiou,
  G.~Quaghebeur, M.A. Rutherford, J.V. Hajnal, J.A. Noble, and J.A. Schnabel.
\newblock {Registration of 3D fetal neurosonography and MRI}.
\newblock \emph{Med Image Anal}, 17:\penalty0 1137--1150, 2013.

\bibitem[Lee et~al.(2019)Lee, Sabuncu, and Dalca]{Lee2019}
H.W. Lee, M.R. Sabuncu, and A.V. Dalca.
\newblock {Few Labeled Atlases are Necessary for Deep-Learning-Based
  Segmentation}.
\newblock 2019.
\newblock URL \url{http://arxiv.org/abs/1908.04466}.

\bibitem[Li et~al.(2020)Li, Zhao, Liu, and Cao]{li2020}
P.~Li, H.~Zhao, P.~Liu, and F.~Cao.
\newblock Automated measurement network for accurate segmentation and parameter
  modification in fetal head ultrasound images.
\newblock \emph{Medical \& Biological Engineering \& Computing}, 58\penalty0
  (11):\penalty0 2879--2892, 2020.

\bibitem[Looney et~al.(2021)Looney, Yin, Collins, Nicolaides, Plasencia,
  Molloholli, Natsis, and Stevenson]{looney2021}
P.~Looney, Y.~Yin, S.L. Collins, K.H. Nicolaides, W.~Plasencia, M.~Molloholli,
  S.~Natsis, and G.N. Stevenson.
\newblock Fully automated 3-d ultrasound segmentation of the placenta, amniotic
  fluid, and fetus for early pregnancy assessment.
\newblock \emph{IEEE Transactions on Ultrasonics, Ferroelectrics, and Frequency
  Control}, 68\penalty0 (6):\penalty0 2038--2047, 2021.

\bibitem[Namburete et~al.(2018)Namburete, Xie, Yaqub, Zisserman, and
  Noble]{Namburete2018}
A.I.L. Namburete, W.~Xie, M.~Yaqub, A.~Zisserman, and J.A. Noble.
\newblock {Fully-automated alignment of 3D fetal brain ultrasound to a
  canonical reference space using multi-task learning}.
\newblock \emph{Med Image Anal}, 46:\penalty0 1--14, 2018.

\bibitem[Oishi et~al.(2019)Oishi, Chang, and Huang]{OISHI2019}
K.~Oishi, L.~Chang, and H.~Huang.
\newblock Baby brain atlases.
\newblock \emph{NeuroImage}, 185:\penalty0 865--880, 2019.

\bibitem[Paladini et~al.(2021)Paladini, Malinger, Birnbaum, Monteagudo, Pilu,
  Salomon, and Timor-Tritsch]{ISUOG2021}
D.~Paladini, G.~Malinger, R.~Birnbaum, A.~Monteagudo, G.~Pilu, L.~J. Salomon,
  and I.~E. Timor-Tritsch.
\newblock Isuog practice guidelines (updated): sonographic examination of the
  fetal central nervous system. part 2: performance of targeted
  neurosonography.
\newblock \emph{Ultrasound in Obstetrics \& Gynecology}, 57\penalty0
  (4):\penalty0 661--671, 2021.
\newblock \doi{https://doi.org/10.1002/uog.23616}.
\newblock URL
  \url{https://obgyn.onlinelibrary.wiley.com/doi/abs/10.1002/uog.23616}.

\bibitem[Rousian et~al.(2009)Rousian, Verwoerd-Dikkeboom, Koning, Hop, Van~der
  Spek, Exalto, and Steegers]{rousian2009}
M.~Rousian, C.M. Verwoerd-Dikkeboom, A.H.J. Koning, W.C. Hop, P.J. Van~der
  Spek, N.~Exalto, and E.A.P. Steegers.
\newblock Early pregnancy volume measurements: validation of ultrasound
  techniques and new perspectives.
\newblock \emph{BJOG: An International Journal of Obstetrics \& Gynaecology},
  116\penalty0 (2):\penalty0 278--285, 2009.

\bibitem[Rousian et~al.(2010)Rousian, Koning, Van~Oppenraaij, Hop,
  Verwoerd-Dikkeboom, Van Der~Spek, Exalto, and Steegers]{rousian2010}
M.~Rousian, A.H.J. Koning, R.H.F. Van~Oppenraaij, W.C. Hop, C.M.
  Verwoerd-Dikkeboom, P.J. Van Der~Spek, N.~Exalto, and E.A.P. Steegers.
\newblock An innovative virtual reality technique for automated human embryonic
  volume measurements.
\newblock \emph{Human reproduction}, 25\penalty0 (9):\penalty0 2210--2216,
  2010.

\bibitem[Rousian et~al.(2013)Rousian, Hop, Koning, {Van Der Spek}, Exalto, and
  Steegers]{Rousian2013}
M.~Rousian, W.~C. Hop, A.~H.J. Koning, P.~J. {Van Der Spek}, N.~Exalto, and
  E.~A.P. Steegers.
\newblock {First trimester brain ventricle fluid and embryonic volumes measured
  by three-dimensional ultrasound with the use of I-Space virtual reality}.
\newblock \emph{Hum Reprod}, 28:\penalty0 1181--9, 2013.

\bibitem[Rousian et~al.(2018)Rousian, Koning, Koster, Steegers, Mulders, and
  Steegers-Theunissen]{Rousian2018}
M.~Rousian, A.H.J. Koning, M.P.H. Koster, E.A.P. Steegers, A.G.M.G.J. Mulders,
  and R.P.M. Steegers-Theunissen.
\newblock {Virtual reality imaging techniques in the study of embryonic and
  early placental health}.
\newblock \emph{Placenta}, 64:\penalty0 S29--S35, 2018.

\bibitem[Rousian et~al.(2021)Rousian, Schoenmakers, Eggink, Gootjes, Koning,
  Koster, Mulders, Baart, Reiss, Laven, Steegers, and
  Steegers-Theunissen]{Rousian2021}
M.~Rousian, S.~Schoenmakers, A.J. Eggink, D.V. Gootjes, A.H.J. Koning, M.P.H.
  Koster, A.G.M.G.J. Mulders, E.B. Baart, I.K.M. Reiss, J.S.E. Laven, E.A.P.
  Steegers, and R.P.M. Steegers-Theunissen.
\newblock {Cohort Profile Update: the Rotterdam Periconceptional Cohort and
  embryonic and fetal measurements using 3D ultrasound and virtual reality
  techniques}.
\newblock \emph{Int J Epidemiol}, pages 1--14, 2021.

\bibitem[Ryou et~al.(2016)Ryou, Yaqub, Cavallaro, Roseman, Papageorghiou, and
  Noble]{Ryou2016}
H.~Ryou, M.~Yaqub, A.~Cavallaro, F.~Roseman, A.~Papageorghiou, and J.A. Noble.
\newblock Automated 3d ultrasound biometry planes extraction for first
  trimester fetal assessment.
\newblock pages 196 -- 204, 2016.

\bibitem[Salomon et~al.(2013)Salomon, Alfirevic, Bilardo, Chalouhi, Ghi, Kagan,
  Lau, Papageorghiou, Raine-Fenning, Stirnemann, Suresh, Tabor, Timor-Tritsch,
  Toi, and Yeo]{ISUOG}
L.J. Salomon, Z.~Alfirevic, C.M. Bilardo, G.E. Chalouhi, T.~Ghi, K.O. Kagan,
  T.K. Lau, A.T. Papageorghiou, N.J. Raine-Fenning, J.~Stirnemann, S.~Suresh,
  A.~Tabor, I.E. Timor-Tritsch, A.~Toi, and G.~Yeo.
\newblock Isuog practice guidelines: performance of first-trimester fetal
  ultrasound scan.
\newblock \emph{Ultrasound Obstet Gynecol}, 41\penalty0 (1):\penalty0 102--113,
  2013.

\bibitem[Steegers-Theunissen et~al.(2016)Steegers-Theunissen,
  Verheijden-Paulissen, van Uitert~E.M., Exalto, A.H., A.J., J.J., J.S.,
  Tibboel, Reiss, and Steegers]{Steegers-Theunissen2016}
R.P. Steegers-Theunissen, J.J. Verheijden-Paulissen, M.F. van Uitert~E.M.,
  Wildhagen, N.~Exalto, Koning A.H., Eggink A.J., Duvekot J.J., Laven J.S.,
  D.~Tibboel, I.~Reiss, and E.A. Steegers.
\newblock Cohort profile: The rotterdam periconceptional cohort (predict
  study).
\newblock \emph{Int J Epidemiol}, 45:\penalty0 374--381, 2016.

\bibitem[Steegers-Theunissen et~al.(2013)Steegers-Theunissen, Twigt, Pestinger,
  and Sinclair]{SteegersTheunissen2013}
R.P.M. Steegers-Theunissen, J.~Twigt, V.~Pestinger, and K.D. Sinclair.
\newblock {The periconceptional period, reproduction and long-term health of
  offspring: the importance of one-carbon metabolism}.
\newblock \emph{Hum Reprod Update}, 19\penalty0 (6):\penalty0 640--655, 2013.

\bibitem[Torrents-Barrena et~al.(2019)Torrents-Barrena, Piella, Masoller, and
  Gratac{\'{o}}s]{Torrents-barrena2019}
J.~Torrents-Barrena, G.~Piella, N.~Masoller, and E.~Gratac{\'{o}}s.
\newblock {Segmentation and classification in MRI and US fetal imaging : Recent
  trends and future prospects}.
\newblock \emph{Med Image Anal}, 51:\penalty0 61--88, 2019.

\bibitem[van~den Heuvel et~al.(2018)van~den Heuvel, de~Bruijn, de~Korte, and
  Ginneken]{heuvel2018}
T.L.A. van~den Heuvel, D.~de~Bruijn, C.L. de~Korte, and B.~van Ginneken.
\newblock Automated measurement of fetal head circumference using 2d ultrasound
  images.
\newblock \emph{PloS one}, 13\penalty0 (8):\penalty0 e0200412, 2018.

\bibitem[Yang et~al.(2018)Yang, Yu, Li, Wen, Luo, Bian, Qin, Ni, and
  Heng]{yang2018}
X.~Yang, L.~Yu, Sh. Li, H.~Wen, D.~Luo, C.~Bian, J.~Qin, D.~Ni, and P.~Heng.
\newblock Towards automated semantic segmentation in prenatal volumetric
  ultrasound.
\newblock \emph{IEEE transactions on medical imaging}, 38\penalty0
  (1):\penalty0 180--193, 2018.

\bibitem[Yaqub et~al.(2013)Yaqub, Cuingnet, Napolitano, Roundhill,
  Papageorghiou, Ardon, and Noble]{Yaqub2013}
M.~Yaqub, R.~Cuingnet, R.~Napolitano, D.~Roundhill, A.~Papageorghiou, R.~Ardon,
  and J.A. Noble.
\newblock Volumetric segmentation of key fetal brain structures in 3d
  ultrasound.
\newblock \emph{International Workshop on Machine Learning in Medical Imaging},
  pages 25 -- 32, 2013.

\bibitem[Yaqub et~al.(2015)Yaqub, Kelly, Papageorghiou, and Noble]{Yaqub2015}
M.~Yaqub, B.~Kelly, A.~T. Papageorghiou, and J.~A. Noble.
\newblock Guided random forests for identification of key fetal anatomy and
  image categorization in ultrasound scans.
\newblock pages 687 -- 694, 2015.

\bibitem[Zhengyang et~al.(2019)Zhengyang, Xu, Zhenlin, and
  Niethammer]{Shen2019}
S.~Zhengyang, H.~Xu, X.~Zhenlin, and M.~Niethammer.
\newblock Networks for joint affine and non-parametric image registration.
\newblock 2019.
\newblock URL \url{http://arxiv.org/abs/1903.08811}.

\end{thebibliography}

\end{document}